\documentstyle[preprint,aps,eqsecnum,tighten]{revtex}
\def\beq{\begin{equation}}
\def\eeq{\end{equation}} 
\def\beqa{\begin{eqnarray}} 
\def\eeqa{\end{eqnarray}} 
\def\bfig{\begin{figure}\vspace{5mm}}

\def\efig{\end{figure}}

\def\bqu{\begin{quote}}
\def\equ{\end{quote}}
\def\bitem{\begin{itemize}}
\def\eitem{\end{itemize}}
\def\ben{\begin{enumerate}}
\def\een{\end{enumerate}}
\def\lsim{\stackrel{<}{\sim}}

\def\order{{\cal O}}
\def\lab{\label}
\def\<{\langle}
\def\>{\rangle}

\begin{document}
\draft
\preprint{}
\title{{\bf A Note on (Spin-)Coherent-State Path Integral}}
\author{Junya Shibata \footnote{email: shibata@cmpt01.phys.tohoku.ac.jp} and Shin Takagi \footnote{email: takagi@cmpt01.phys.tohoku.ac.jp}}
\address{Department of Physics, Tohoku University, Sendai 980-8578, Japan}
\maketitle

\bigskip
\begin{abstract}
It is pointed out that there are some fundamental difficulties 
with the frequently used continuous-time formalism 
of the spin-coherent-state path integral.
They arise already in a single-spin system 
and at the level of the "classical action" 
not to speak of fluctuations around the "classical path". 
Similar difficulties turn out to be present in the case 
of the (boson-)coherent-state path integral as well;
although partially circumventable by an ingenious trick
(Klauder's $\epsilon$-prescription) at the "classical level",
they manifest themselves at the level of fluctuations.
Detailed analysis of the origin of these difficulties makes it clear
that the only way of avoiding them is to work with 
the proper discrete-time formalism.
The thesis is explicitly illustrated with a harmonic oscillator 
and a spin under a constant magnetic field.     
\end{abstract}
\pacs{}



\newpage

\widetext
\section{Introduction}

Path integrals are widely used in various fields of physics \cite{{Feynman},{Schulman},{Kleinert},{Swanson}}.
They are thought to be particularly suited 
to semi-classsical evaluation of quantum mechanical transition amplitudes 
(or partition functions), since apparently they can incorporate 
a classical picture much more easily than the operator formalism can.
In particular many works on spin systems make use of 
the spin-coherent-state (i.e., the SU(2)-coherent-state) path integral.
To list just a few of the notable applications,
precession of a single spin under a constant magnetic field \cite{Kuratsuji},
one dimensional anti-ferromagnets \cite{Fradkin},
tunneling of a giant spin in a mesoscopic magnet \cite{Chudnovsky},
tunneling of a magnetic domain wall \cite{Braun-Loss},
and so on. 
The standard starting point, in the case of a single spin of magnitude $S$,
is the following expression for the transition amplitude:
\begin{mathletters}
\beqa
&& \< {\bf n}_F| e^{-i {\hat H} T / \hbar} |{\bf n}_I \> = \int_{SCS} {\cal D}\theta {\cal D}\phi \exp \left( \frac{i}{\hbar}\tilde{S}_{SCS}[\theta,\phi] \right),  \label{ctscspi}\\
&& \tilde{S}_{SCS}[\theta,\phi]:=\int_{0}^{T}dt \left\{ \hbar S ( \cos \theta(t) - 1 )\dot{\phi}(t)  - H( \theta(t),\phi(t) ) \right\}. \lab{ctscspi-action}
\eeqa\label{ctscs}\end{mathletters}(Throughout the paper, the equation A:=B denotes that A is defined by B.) Here, $|{\bf n}_{\alpha}\>$ with ${\bf n}_{\alpha}$ being a unit vector ($\alpha =$ {\it I} or {\it F} ) is 
the spin coherent state in which the spin may be visualized as 
oriented along ${\bf n}_{\alpha}$.
In the integral, $\theta$ and $\phi$ denote polar and azimuthal angles of 
the spin orientation at intermediate times,
and $H(\theta,\phi)$ is the Hamiltonian in the spin-coherent-state representation.
(The precise definition of various symbols is given in Sec. II.)
Time $t$ is treated as a continuous parameter in the above expression,
which therefore may be called 
the {\em continuous-time spin-coherent-state path integral},
to be abbreviated as CTSCSPI.  

As we will see, the expression leads to a grave difficulty.
If one tries to evaluate it in the spirit of semi-classical approximation,
one fails already at the "classical level". 
If one ignores this failure and proceeds to integrate over fluctuations,
one obtains a meaningless result.
In order to appreciate the difficulty, 
it is worthwhile to recall the coherent-state 
(i.e., boson-coherent-state) path integral,
which is in a sense a linear version of the spin-coherent-state path integral.
In the case of a single particle governed by a Hamiltonian $H(p,q)$,
the relevant transition amplitude is often expressed as 
\begin{mathletters}
 \beqa
&& \< p_F, q_F| e^{-i {\hat H} T / \hbar} |p_I ,q_I \> = \int_{CS} {\cal D}p {\cal D}q 
\exp \left(  \frac{i}{\hbar} \tilde{S}_{CS}[p,q]  \right), 
 \label{ctcspi} \\
&& \tilde{S}_{CS}[p,q]:=\int_{0}^{T}dt \left\{ \frac{1}{2}( p(t)\dot{q}(t) - \dot{p} (t)q(t) )  - H( p(t),q(t) ) \right\}, 
 \label{ctcspi-action}
\eeqa\label{ctcs}\end{mathletters}where $| p_{\alpha} ,q_{\alpha} \>$ ($\alpha =$ {\it I} or {\it F} ) is the coherent state
labeled by the complex number $q_{\alpha} + i p_{\alpha}$.
This expression may be called 
the {\em continuous-time coherent-state path integral},
to be abbreviated as CTCSPI. 
It will be seen that this innocent-looking expression also contains
difficulties.
In order to appreciate them, it is in turn worthwhile to compare (\ref{ctcs})
with the phase-space path integral expression for a Feynman kernel.
In the case of the single particle, 
it is often expressed as
\begin{mathletters}
 \beqa
&& \< q_{F}|e^{-i\hat{H}T/\hbar}| q_{I} \> 
=
\int_{PS} {\cal D}p {\cal D}q \exp \left( \frac{i}{\hbar} S_{PS}[p,q] \right),
\label{ctpspi} \\
&& S_{PS}[p,q]:= \int_{0}^{T}dt \{ p(t)\dot{q}(t) - H( p(t),q(t) ) \},
\lab{ctps-action}
 \eeqa
\label{ctps}\end{mathletters}where $|q_{\alpha}\>$ ($\alpha =$ {\it I} or {\it F} ) is a position eigenket.
This expression may be called 
the {\em continuous-time phase-space path integral},
to be abbreviated as CTPSPI.

For the understanding of the announced difficulties 
associated with (\ref{ctscs}) and (\ref{ctcs}),
it is of vital importance to appreciate the difference between (\ref{ctcs}) and (\ref{ctps})
in spite of their apparent similarity.
Therefore, we will begin by reviewing the semiclassical evaluation
of (\ref{ctps}), which will be followed by that of (\ref{ctcs}) and (\ref{ctscs}).
In the latter two cases, in contrast to the case of CTPSPI, 
one can not in general find a "classical path".
Even if one somehow circumvent this difficulty,
one obtains a wrong value for the "classical action".
Fluctuation integrals lead to further non-sensical result.
In Sec. IV, we critically review what is to be called Klauder's $\epsilon$-prescription \cite{Klauder},
and point out that 
ambiguities arise in dealing with fluctuation especially in the case of the spin-coherent-state path integral. 
Section V is devoted to a thorough re-examination of the whole issue
in the proper discrete-time formalism. 
First, it is shown that the difficulties concerning 
the "classical path" and "classical action" disappear;
they are shown to be illusions caused by the ambiguity 
of the continuous-time formalism.
Second, it is clarified how the $\epsilon$-prescription
is related to the discrete-time formalism. 
We present these discussions in a concrete form by explicitly working out 
the examples of a harmonic oscillator for the coherent-state case 
and a spin under a constant magnetic field for the spin-coherent-state case
both in the continuous-time and discrete-time formalism.

Our conclusion, then, is that any serious work with 
(spin-)coherent-state path integrals should be based on 
their discrete-time form.

Part of the difficulties associated with CTSCSPI was 
previously noted by Funahashi et al \cite{{Funahashi1},{Funahashi2}} and by Schilling \cite{Schilling}. The discrete-time formalism was employed by Solari \cite{Solari} 
who developed a general method of evaluating the fluctuation integral,
and by Funahashi et al who evaluated the partition function 
for a single spin under a constant magnetic field
\cite{fn1}.
However, the nature of the difficulties associated with CTSCSPI
seems to have been left unscrutinized; 
many workers still use CTSCSPI or its Klauder-augmented version
(to be calle KCTSCSPI) because of their apparent simplicity.
We hope that the present paper serves to warn the users of 
the spin-coherent-state path integral against uncritical use of
 CTSCSPI or KCTSCSPI.
    
\section{Notation}
In the case of the phase-space and coherent-state path integrals,
we consider a single-particle system.
The position and momentum of the particle are denoted by $q$ and $p$
which are measured in units such that both of them have dimension of 
$\hbar^{1/2}$.
The corresponding operators are marked by a caret.
Accordingly
\beq
[\hat{q},\hat{p} ] = i \hbar.
\eeq
We introduce
\beq
\hat{a} := \sqrt{\frac{1}{2\hbar}}( \hat{q} + i\hat{p} ),\qquad  \hat{a}^{\dagger} := \sqrt{\frac{1}{2\hbar}}( \hat{q} - i\hat{p} ),
\label{a-q-p}
\eeq
and their c-number counterparts:
\beq
\xi := \sqrt{\frac{1}{2\hbar}}( q + ip ) ,\qquad  \xi^* := \sqrt{\frac{1}{2\hbar}}( q - ip ) .
\label{xi-q-p}
\eeq
It is to be understood that $\xi_\alpha$ and $ \xi^*_\alpha$ are
related to $q_{\alpha}$ and $p_{\alpha}$ 
in the above fashion for any subscript $\alpha$. 
The coherent state is defined in the standard way as
\beq
 |p,q \> \equiv |\xi \> := \exp \left[ \xi \hat{a}^{\dagger} - \xi^{*} \hat{a}\right] | 0  \> = \exp \left[ \frac{i}{\hbar} ( p \hat{q} - q \hat{p} ) \right] | 0 \>,
\label{coherent-state} 
\eeq
with
\beq
\hat{a} |0\> = 0, \qquad \<0|0\> = 1.
\eeq
Hence
\begin{mathletters}
\beqa
&& |\xi\> = e^{-|\xi|^2/2} e^{\xi \hat{a}^\dagger} |0\>, \qquad \hat{a} | \xi\> = \xi |\xi\>, \\
&& \<\xi | \xi'\> = \exp \left[ -\frac{1}{2}( |\xi|^{2} + |\xi'|^{2} ) + \xi^{*}\xi' \right].
\label{cs-innerp}
\eeqa 
\end{mathletters}For an illustration we treat the harmonic oscillator
governed by the Hamiltonian
\beq
\hat{H} \equiv H (\hat{p}, \hat{q}) := \frac{1}{2}(\hat{p}^2 + \hat{q}^2 - \hbar) 
 = \hbar \hat{a}^{\dagger} \hat{a}.
\label{cs-ho-hamiltonian}
\eeq
(By convention $H$ has the dimension of $\hbar$, and time is dimensionless.)
Under this Hamiltonian the coherent state evolves in time as
\beq
e^{-i\hat{H}T/\hbar} | \xi \> = | \xi e^{-iT} \>.
\eeq
It follows that
\beqa
 \<p_F,q_F| e^{-i {\hat H} T / \hbar} |p_I, q_I\>
 &=& \<\xi_F|e^{-i {\hat H} T / \hbar}|\xi_I\> \nonumber \\
 &=& \< \xi_{F} | \xi_{I} e^{-iT} \> 
 = \exp \left[ -\frac{1}{2} (  | \xi_{F} |^{2} + | \xi_{I} |^{2} ) + \xi^{*}_{F} \xi_{I} e^{-iT} \right].
\label{exact-cs-ta}
\eeqa
The matrix element of ${\hat H}$ in the coherent-state representation
is given by 
\beq
 {\cal H}(\xi^*, \xi') := \frac{\<\xi|{\hat H} |\xi'\>}{\<\xi| \xi'\>}
  = \hbar \xi^* \xi'.
\label{cs-hamiltonian}
\eeq
This is a function of $\xi^*$ and  $\xi'$ alone and 
involves neither $\xi$ nor $(\xi')^*$.
(This property holds not only for a harmonic oscillator but also 
for any system.)
 
In the case of the spin-coherent-state path integral,
we consider a system of a single spin of magnitude $S$.
The dimensionless spin operator is denoted by ${\hat {\bf S}}$,
whose components obey 
\beq
[\hat{S}_x, \hat{S}_y ] = i \hat{S}_z,  \qquad {\rm and ~cyclic}.
\eeq
We introduce an auxiliary unit vector ${\bf n}$ whose
polar and azimuthal angles are $\theta$ and $\phi$,
respectively, 
and also a complex number $\xi$ corresponding to 
the Riemann projection of ${\bf n}$:
\beq
 \xi := e^{i \phi} \tan \frac{\theta}{2}, \qquad  \xi^* := e^{-i \phi} \tan \frac{\theta}{2} .
\label{xi-phi-theta}
\eeq
It is to be understood that $\xi_\alpha$ and $ \xi^*_\alpha$ are
related to $\phi_\alpha$ and $\theta_\alpha$ 
in the above fashion for any subscript $\alpha$. 
 The spin coherent state \cite{Radcliffe} is defined as
\beqa
 |{\bf n}\> \equiv |\xi\> &:=&  \exp \left[- \zeta^{*} \hat{S}_{+} + \zeta \hat{S}_{-}  \right] | S \>   \nonumber \\
                          &=& ( 1 + |\xi|^2 )^{-S} \sum_{M=-S}^{S}\left(\frac{(2S)!}{(S-M)!(S+M)!}\right)^{1/2} \xi^{S-M} | M \>, \qquad \zeta := e^{i\phi}\frac{\theta}{2}, 
\label{spin-coherent-state}
\eeqa
with
\beq
 \hat{S}_\pm = \hat{S}_x \pm i\hat{S}_y  , \qquad \hat{S}_z |S\> = S |S\> , \qquad \<S| S\> = 1.
\eeq
Hence
\begin{mathletters}\beqa
&& {\bf n} \cdot {\hat {\bf S}} |\xi\> = S |\xi\>,
\\
&& \<\xi|\xi'\> =  \frac{(1 + \xi^{*}\xi')^{2S}}{(1 + |\xi|^{2})^{S}( 1 + |\xi'|^{2})^{S}}.
\label{scs-innerp}
\eeqa 
\end{mathletters}For an illustration we treat the system
governed by the Hamiltonian
\beq
\hat{H} :=  - \hbar \hat{S}_z
\label{scs-constmag-hamiltonian}
\eeq
which represents a spin under a constant magnetic field.
It is useful to note that
\beq
  e^{i \hat{S}_z T} \hat{S}_\pm e^{- i \hat{S}_{z} T} = e^{ \pm i T} \hat{S}_\pm.\eeq
Therefore, under the Hamiltonian (\ref{scs-constmag-hamiltonian}), the spin coherent state evolves in time as
\beq
   e^{-i \hat{H} T} |\xi\> = 
   \left. |\xi\> \right|_{\zeta \rightarrow \zeta e^{- i T}} = e^{iST}|\xi e^{-iT} \>.
\eeq
It follows that
\beqa
 \<{\bf n}_F| e^{-i {\hat H} T / \hbar} |{\bf n}_I\>
 &=& \<\xi_F| e^{-i {\hat H} T / \hbar} |\xi_I\> \nonumber \\
 &=& \frac{( 1 + \xi^{*}_{F}\xi_{I}e^{-iT})^{2S}}{(1 + |\xi_{F}|^{2})^{S}(1 + |\xi_{I}|^{2})^{S}}e^{iST}.
\label{exact-scs-ta}
\eeqa
The matrix element of ${\hat H}$ in the spin-coherent-state representation
is given by 
\beq
 {\cal H}(\xi^*, \xi') := \frac{\<\xi|{\hat H} |\xi'\>}{\<\xi| \xi'\>}
  = - \hbar S \frac{1 - \xi^* \xi'}{1 + \xi^* \xi'}.
\label{scs-hamiltonian}
\eeq
The remark made on Eq.~(\ref{cs-hamiltonian}) applies to this equation as well.
       
\section{Continuous-Time Formalism}
In this section we discuss stationary-action approximations for 
continuous-time path integrals. 
\subsection{Continuous-Time Phase-Space Path Integral}
One would expect that CTPSPI (\ref{ctps}) is dominated 
by the statinary-action path $(p^{cl}(t),q^{cl}(t))$
at which the action $S_{PS}[p,q]$ is stationary.
The stationary-action path is determined by
\begin{mathletters}
 \beqa 
&& 0 = \left. \frac{\delta S_{PS}[p,q]}{\delta p(t)} \right|_{cl} 
= \dot{q}^{cl}(t) - \left. \frac{\partial H(p,q)}{\partial p} \right|_{cl},
\label{eqmo-ctps-sap1}
\\
&& 0 = \left. \frac{\delta S_{PS}[p,q]}{\delta q(t)} \right|_{cl} 
= -\dot{p}^{cl}(t) - \left. \frac{\partial H(p,q)}{\partial q} \right|_{cl},
\label{eqmo-ctps-sap2}
 \eeqa\label{eqmo-ctps-sap}\end{mathletters}where the symbol $|_{cl}$ indicates the replacement 
$(p,q) \rightarrow (p^{cl}(t),q^{cl}(t))$ after differentiation.
On inspection of the left-hand-side of (\ref{ctpspi}), one would argue that 
the space of paths to be integrated over is defined by 
the "boundary condition"
\beq
q(T) = q_{F}, \qquad q(0) = q_{I},
\label{bc-ctps1}
\eeq
and accordingly that the same condition should be imposed on the purported dominant path:
\beq 
q^{cl}(T) = q_{F}, \qquad q^{cl}(0)=q_{I}.
\label{bc-ctps2}
\eeq
Being a couple of first-order differential equations, 
the above set of equations has a solution under this boundary condition
\cite{fn2}.
Note that no boundary condition is imposed on $p^{cl}$; 
the values of $p^{cl}(T)$ and $p^{cl}(0)$ are determined 
{\em a posteriori}.
Obviously the stationary-action path is a solution of 
the Hamilton equation of motion and deserves the name
"classical path"; hence the superscript $cl$.
Once a classical path is found, 
one may decompose $(p(t),q(t))$ at intermediate times into 
a sum of $(p^{cl}(t),q^{cl}(t))$ and fluctuations around it,
thereby proceeding to integration over the fluctuations.
The rest is a well-known story and need not be repeated here.
The result so obtained is known to be correct.
(One might as well note a subtle point which is often ignored;
the claim that the pahse-space path integral is dominated 
by the classical path as defined above is not
correct. See Sec. V.)
Since $S_{PS}[p,q]$ remains real throughout the calculation,
the above procedure may also be called a stationary-phase approximation.

\subsection{Continuous-Time Coherent-State Path Integral}
Let us review CTCSPI in parallel with the previous subsection.
Let the stationary-action path be $(p^S(t),q^S(t))$,
which is determined by
\begin{mathletters}
 \beqa
&& 0 = \left. \frac{\delta \tilde{S}_{CS}[p,q]}{\delta p(t)} \right|_S 
     = \dot{q}^{S}(t) - \left. \frac{\partial H(p,q)}{\partial p} \right|_{S},
\label{eqmo-ctcs-sap1}
\\
&& 0 = \left. \frac{\delta \tilde{S}_{CS}[p,q]}{\delta q(t)} \right|_S 
     = -\dot{p}^{S}(t) - \left. \frac{\partial H(p,q)}{\partial q} \right|_{S},
\label{eqmo-ctcs-sap2}
\eeqa\label{eqmo-ctcs-sap}\end{mathletters}where the symbol $|_S$ indicates the replacement 
$(p,q) \rightarrow (p^S(t),q^S(t))$ after differentiation.
This set of equations is also identical with the Hamilton equation of motion.
On inspection of the left-hand-side of (\ref{ctcspi}), one would now think that
the space of paths in the present case is defined not by
the boundary condition (\ref{bc-ctps1}) but by 
\beq
p(T) = p_{F},\qquad q(T) = q_{F}, \qquad p(0) = p_{I}, \qquad q(0) = q_{I},
\label{bc-ctcs-qp}
\eeq
and accordingly require 
\beq
p^{S}(T) = p_{F},\qquad q^{S}(T) = q_{F}, \qquad p^{S}(0) = p_{I}, \qquad q^{S}(0) = q_{I}.
\label{bc-ctcs-qp2}
\eeq
However a couple of first-order differential equations can not
in general accomodate a set of four conditions.
This is the first difficulty \cite{Schul}.
A way to evade this difficulty would be to note that
the above boundary condition is motivated by the notation $\<p_F, q_F|$ and $|p_I, q_I\>$, 
which is rather misleading.
Unlike the ket $|q\>$ corresponding to the definite position $q$,
the state $|p,q\>$ does not correspond to a definite "momentum and position".
(In the latter state, both momentum and position have indeterminacy 
of ${\cal O}(\hbar^{1/2})$.)
It is more appropriate to label the state by a single complex number $\xi$
which is related to $(p, q)$ via Eq.~(\ref{xi-q-p}).
Accordingly, one would re-write Eqs.~(\ref{eqmo-ctcs-sap}) in terms of 
\beq
  \xi^S(t) := \sqrt{\frac{1}{2\hbar}}( q^S(t) + i p^S(t))  , \qquad {\rm and} \qquad  {\bar \xi}^S(t) := \sqrt{\frac{1}{2\hbar}}( q^S(t) - i p^S(t)), 
\lab{sap-xi-q-p}
\eeq
where the symbol ${\bar \xi}^S(t)$ is used instead of $\{ \xi^S(t) \}^*$
for the reason to be explained shortly.
Equivalently one may re-express the action in terms of $\xi$ and $\xi^*$
as 
\beqa
 S_{CS}[\xi^*, \xi] 
    &:=& \tilde{S}_{CS}\left[ \sqrt{\frac{\hbar}{2}} \frac{1}{i}(\xi - \xi^*),
              \sqrt{\frac{\hbar}{2}} (\xi + \xi^*)\right]\nonumber\\
    &=& \int_0^T dt \left\{\frac{i\hbar}{2}( \xi^*(t) \dot{\xi}(t) - \dot{\xi}^{*}(t) \xi(t) ) - {\cal H}(\xi^{*}(t) , \xi(t) )  \right\},
\label{action-ctcs-xi}
\eeqa
and vary it with respect to $\xi$ and $\xi^*$
formally regarding them as mutually independent. 
One would then find
\beq
 \dot{\xi}^{S}(t)       =  -\left.\frac{i}{\hbar}\frac{\partial {\cal H}(\xi^{*},\xi)}{\partial \xi^{*}}\right|_{S} =  -i \xi^{S}(t), \qquad 
 \dot{\bar{\xi}^{S}}(t) = \left.\frac{i}{\hbar}\frac{\partial {\cal H}(\xi^{*},\xi)}{\partial \xi}\right|_{S} = i \bar{\xi}^{S}(t),
\label{eqmo-ho-ctcs-xi}
\eeq
where the symbol $|_{S}$ indicates the replacement $(\xi^{*},\xi) \to (\bar{\xi}^{S}(t),\xi^{S}(t))$ after differentiation, and the last equalities hold for the harmonic oscillator.
Now one would make a crucial observation; 
since the normalization factor can be taken care of separately,
the state $|\xi_\alpha \>$ ($\alpha = {\it I}$ or ${\it F}$) may be replaced by
\beq
 |\xi_\alpha) := e^{|\xi_\alpha|^2/2} |\xi_\alpha \>
     = e^{\xi_\alpha {\hat a}^\dagger}|0 \>.
\label{nonnorm-cs}
\eeq
Then the amplitude $(\xi_F| e^{-i\hat{H}T/\hbar}|\xi_I)$ does not depend 
on $\xi_F$ nor on $\xi_I^*$ but depends only on $\xi_F^*$ and $\xi_I$.
One could thus argue that
the relevant space of paths is defined by the boundary condition 
\cite{Itzykson}
\beq
\xi^{*}(T) = \xi^{*}_{F},\qquad \xi(0) = \xi_{I},
\lab{bc-ctcs-xi2}
\eeq
and that the boundary condition to be imposed on
the set of Eqs.~(\ref{eqmo-ho-ctcs-xi}) is
\beq
\bar{\xi}^{S}(T) = \xi^{*}_{F}, \qquad \xi^{S}(0) = \xi_{I}.
\label{bc-ctcs-xi}
\eeq
With this boundary condition, the equations can be solved to yield
\beq
\xi^{S}(t) = \xi_{I} e^{-it}, \qquad \bar{\xi}^{S}(t) = \xi^{*}_{F} e^{ i( t - T )}. \lab{sol-ho-ctcs}
\eeq
The price to be paid is that ${\bar \xi}^S(t)$ is 
in general different from the complex conjugate of $\xi^S(t)$. 
This is the reason of the notation used. 
As a result, $(p^S(t),q^S(t))$ related to $(\xi^S(t), {\bar \xi}^S(t))$ 
via (\ref{sap-xi-q-p}) are not in general real. 
However, they are real if and only if 
\beq 
 \xi_F = \xi_I e^{-iT},\label{special-bc}
\eeq  
in which case they describe a classical path,
that is, a real solution of the Hamilton equation of motion.
The appearance of a complex stationary-action path does not 
by itself cause any difficulty;
contours of integration over each $p(t)$ and $q(t)$ at intermdediate times,
which are originally defined to be along the real axis, 
may be distorted into the respective complex plane 
so that they as a whole constitute a steepest-descent contour 
through the saddle point $(p^S,q^S)$.
The steepest-descent method of course entails the decomposition 
of $p(t)$ and $q(t)$ into a sum of the stationary-action path 
and fluctuations around it.
In this procedure, the action $\tilde{S}_{CS}[p,q]$ does not remain real
but become complex in general.
Hence it is inappropriate 
to call the procedure a stationary-phase approximation.
By the same token it is misleading to call 
the statioanry-action path a classical path.
(In passing, note that Eqs.~(\ref{eqmo-ho-ctcs-xi}) 
with the boundary condition (\ref{bc-ctcs-xi}) has a solution for arbitrary $T$
in contrast to the case of Eqs.~(\ref{eqmo-ctps-sap}) with the boundary condition (\ref{bc-ctps2}).)

One might thus hope that the CTCSPI could be worked out 
by imposing the boundary condition (\ref{bc-ctcs-xi}).
Unfortunately the action $S_{CS}^{SAP}$ associated with the stationary-action path 
\beqa
  S_{CS}^{SAP} &:=& S_{CS}[\bar{\xi}^{S},\xi^{S}]\nonumber\\ 
               &=& \int_{0}^{T} \left\{ \frac{1}{2} \left(\bar{\xi^{S}}(t)\left.\frac{\partial {\cal H}(\xi^{*},\xi)}{\partial \xi^{*}}\right|_{S} + \left.\frac{\partial {\cal H}(\xi^{*},\xi)}{\partial \xi}\right|_{S}\xi^{S}(t) \right) -  {\cal H}(\bar{\xi}^{S}(t),\xi^{S}(t) ) \right\}
\eeqa
vanishes in the case of the harmonic oscillator for which ${\cal H}$ is bilinear in $\xi^{*}$ and $\xi$.
This is the second difficulty, 
since one would have hoped that $S_{CS}^{SAP}$ agrees 
with the exponent of (\ref{exact-cs-ta}); 
in the quasi-classical situation where $p_\alpha$ and $q_\alpha$ are regarded 
as of ${\cal O}(\hbar^0)$, the exponent is of ${\cal O}(\hbar^{-1})$ .
Suppose one disregarded this difficulty and proceeded to 
make the following replacement in (\ref{ctcs}):
\beq
\xi(t) = \xi^{S}(t) + \eta(t) , \qquad \xi^{*}(t) = \bar{\xi}^{S}(t) + \eta^{*}(t) ,\label{expand-xiseta}
\eeq
where $\eta^*(t)$ is the complex conjugate of $\eta(t)$ \cite{independent}. 
Since the boundary condition has been taken care of by 
the stationary-action path, one would restrict $\eta(t)$ so that
\beq
\eta^{*}(T) = 0, \qquad \eta(0) = 0.
\label{bc-ctcs-fluc-eta}
\eeq
By definition of the stationary-action path, 
the action does not contain terms linear in the fluctuations $\eta(t)$.
It takes the form
\begin{mathletters}
\beqa
&&S_{CS}[ \bar{\xi}^{S} + \eta^{*}, \xi^{S} + \eta ] \simeq S_{CS}^{SAP} + S^{(2)}_{CS}[\eta^{*} , \eta] , \label{expand-sapfluc}\\
&& S^{(2)}_{CS}[ \eta^{*} , \eta ] := i\hbar \int_{0}^{T} dt \eta^{*}(t)\left(\frac{d}{dt} + i \right) \eta(t).
\eeqa
\end{mathletters}If one formally integrated over the fluctuations, one would obtain
\beq
\int {\cal D}\eta {\cal D}\eta^{*} 
\exp\left( \frac{i}{\hbar} S^{(2)}_{CS}[ \eta^{*} , \eta] \right) \propto  \left( \det\left( \frac{d}{dt} + i  \right) \right)^{-1} ,
\label{int-ctcs-fluc}
\eeq
where the formal determinant denotes the product 
of the eigenvalues of the differential operator $ d/dt + i$, 
which is supposed to act on the space of functions 
satisfying the Dirichlet boundary condition (\ref{bc-ctcs-fluc-eta}).
Obviously, this differential operator does not possess an eigenfunction. 
Hence the formal determinant does not exist.
This is the third difficulty.

\subsection{Continuous-Time Spin-Coherent-State Path Integral}
It is easy to see that 
the situation with CTSCSPI is largely the same as in the case of CTCSPI.
Thus, a stationary-action path $(\theta^S(t), \phi^S(t))$ satisfying
the boundary condition
\beq
\theta^{S}(T) = \theta_{F},\qquad \phi^{S}(T) = \phi_{F}, \qquad \theta^{S}(0) = \theta_{I},\qquad \phi^{S}(0) =\phi_{I}   
\label{bc-ctscs-phitheta}
\eeq
does not exist.
Again, the spin coherent state $|\xi\>$,
apart from the normalization factor $(1 + |\xi|^2)^{-S}$,
does not depend on $\xi^*$ but depend only on $\xi$.
Hence one would proceed as follows.
One would re-expresses the action as
\beqa
 S_{SCS}[\xi^*, \xi] 
    &&:= \tilde{S}_{SCS}\left[2 \tan^{-1}(|\xi|), \frac{1}{2i}\ln \left(\frac{\xi}{\xi^{*}}\right)\right]  \nonumber \\
    &&= \int_{0}^{T}dt \left[\frac{i\hbar S}{ 1 + |\xi(t)|^{2}} \left( \xi^{*}(t)\dot{\xi}(t) 
           - \dot{\xi}^{*}(t)\xi(t) \right) - {\cal H} ( \xi^{*}(t) , \xi(t) ) \right] \label{action-ctscs}
\eeqa
and vary it with respect to $\xi$ and $\xi^*$, 
formally regarding them as mutually independent. 
One would then find
\begin{mathletters}\beqa
&& \dot{\xi}^{S}(t) = - \left. \frac{i( 1 + \bar{\xi}^{S}(t)\xi^{S}(t) )^{2}}{2\hbar S} \frac{\partial {\cal H}(\xi^{*} , \xi) }{\partial \xi^{*}}\right|_{S}                                   = -i \xi^{S}(t), \\
&& \dot{\bar{\xi}^{S}}(t) = \left. \frac{i( 1 + \bar{\xi}^{S}(t)\xi^{S}(t) )^{2}}{2\hbar S}  \frac{\partial {\cal H}(\xi^{*} , \xi) }{\partial \xi}\right|_{S}                                       = i \bar{\xi}^{S}(t),
\label{eqmo-constmag-ctscs-xi}
\eeqa
\end{mathletters}where the last equalities hold for the spin under a constant magnetic field.
Arguing that the boundary condition to be imposed is (\ref{bc-ctcs-xi}), 
one would obtain the solution which is formally identical to (\ref{sol-ho-ctcs}).
The action $S_{SCS}^{SAP}$ associated with this stationary-action path 
would then be found as
\beqa
  S_{SCS}^{SAP} &:=& S_{SCS}[ \bar{\xi}^{S}, \xi^{S} ] \nonumber \\
&=&  \int_{0}^{T} dt 
         \left\{ \frac{1}{2}( 1 + \bar{\xi}^{S}(t)\xi^{S}(t)) 
         \left( \bar{\xi}^{S}(t)\left.\frac{\partial {\cal H}(\xi^{*} , \xi) }{\partial \xi^{*}}\right|_{S}
                + \left.\frac{\partial {\cal H}(\xi^{*} , \xi) }{\partial \xi}\right|_{S} \xi^{S}(t) \right) 
              - {\cal H} ( \bar{\xi}^{S}(t) , \xi^{S}(t) )   \right\} 
                                                      \nonumber \\
&=& \hbar ST,  \label{specialcasevalue}
\eeqa 
which does not lead to the desired result (\ref{exact-scs-ta})
except for the special case of (\ref{special-bc}).
Integration over the fluctuations also leads to 
the same sort of difficuly as in CTCSPI.

\section{Klauder's $\epsilon$-Prescription}
Klauder \cite{Klauder} insisted on having a stationary-action path
satisfying the boundary condition (\ref{bc-ctcs-qp2}).
He augmented the action by what is to be called Klauder's $\epsilon$-term.
According to him, it is motivated by the metric of 
the relevant phase space.

\subsection{Klauder's Continuous-Time Coherent-State Path Integral}
In the case of the coherent state,  
the phase space is a plane whose metric is $(dp)^2 + (dq)^2$ or equivalently $2\hbar |d\xi|^{2}$.
Klauder's augmented action, to be denoted by $S_{KCS}[\xi^{*},\xi]$, reads
\begin{mathletters}
 \beqa
 S_{KCS}[\xi^{*},\xi] := S_{CS-\epsilon}[\xi^{*},\xi] + S_{CS}[\xi^{*},\xi] ,
               \label{action-kctcs}    
\eeqa
where
\beqa
  S_{CS-\epsilon}[\xi^{*},\xi] 
    := \int_0^T dt \frac{i\hbar}{2}\epsilon|\dot{\xi}(t)|^{2} 
\label{action-eterm-kctcs}
 \eeqa
  \label{kaluder-action}\end{mathletters}with $\epsilon$ being an infinitesimal positive number.
Accordingly Klauder's stationary-action path, to be denoted by $(\bar{\xi}^{KS}(t), \xi^{KS}(t))$ and specialized to the harmonic oscillator, 
obeys the following set of equations:
\begin{mathletters}
 \beqa
&&  \frac{\epsilon}{2}\ddot{\bar{\xi}}~^{KS}(t) + \dot{\bar{\xi}}~^{KS}(t) - i\bar{\xi}^{KS}(t) = 0  ,
\label{eqmo-kctcs-sap1}
\\
&&  \frac{\epsilon}{2}\ddot{\xi}^{KS}(t) - \dot{\xi}^{KS}(t) - i\xi^{KS}(t) = 0 .    
\label{eqmo-kctcs-sap2}
 \eeqa
 \label{eqmo-kctcs-sap}\end{mathletters} This set of equations, 
being a couple of second-order differential equations, 
can accomodate the boundary condition (\ref{bc-ctcs-qp2}), namely
\beqa
\bar{\xi}^{KS}(T) = \xi^{*}_{F},\qquad \xi^{KS}(T)=\xi_{F},\qquad \bar{\xi}^{KS}(0) = \xi^{*}_{I},\qquad \xi^{KS}(0) =\xi_{I}.
\eeqa
Although these equations can be solved for arbitary $\epsilon$, we may proceed as follows for an infinitesimal $\epsilon$. In Eq. (\ref{eqmo-kctcs-sap1}), the first term is effective only for the initial interval $0 < t \lsim \epsilon$, where it forces $\bar{\xi}^{KS}(t)$ change sharply. Thus it is convenient to put 
\beqa 
\bar{\xi}^{KS}(t) = \bar{\chi}(t)\bar{\xi}^{S}(t) \label{chixis1}
\eeqa
with $\bar{\xi}^{S}(t)$ given by (\ref{sol-ho-ctcs}). Here $\bar{\chi}(t)$ is {\em essentially} unity (i.e., approximately equal to unity with corrections only of $\order(\exp(-T/\epsilon))$) except for the initial interval, where it changes sharply so as to ensure the condition $\bar{\xi}^{KS}(0) = \xi^{*}_{I}$. Hence $\bar{\chi}(t), \dot{\bar{\chi}}(t)$ and $\ddot{\bar{\chi}}(t)$ are at most of $\order(\epsilon ^{0}), \order(\epsilon ^{-1})$ and $\order(\epsilon ^{-2})$, respectively.
Similar consideration applies to Eq. (\ref{eqmo-kctcs-sap2}). Thus, if we put 
\beqa
\xi^{KS}(t) = \chi(t)\xi^{S}(t)  \label{chixis2}
\eeqa
with $\xi^{S}(t)$ given by (\ref{sol-ho-ctcs}), then $\chi(t)$ is essentially unity except for the final interval $T-\epsilon \lsim t < T$, where it changes sharply.
Up to $\order(\epsilon^{0})$, Eq. (\ref{eqmo-kctcs-sap}) takes the form
\beqa
&&\frac{\epsilon}{2}\ddot{\bar{\chi}}(t) + ( 1 + i\epsilon)\dot{\bar{\chi}}(t) = 0 , \\
&&\frac{\epsilon}{2}\ddot{\chi}(t) - ( 1 + i\epsilon)\dot{\chi}(t) = 0 .
\eeqa
This may be solved to yield 
\beqa
&&\bar{\chi}(t) = 1 + (\bar{\chi}(0)-1)e^{-  \lambda t} + \order(\epsilon) , \\
&&\chi(t) = 1 + (\chi(T) - 1)e^{- \lambda(T-t)} + \order(\epsilon),
\eeqa
where
\beqa
\bar{\chi}(0) = \xi^{*}_{I}/\bar{\xi}^{S}(0), \qquad \chi(T) = \xi_{F}/\xi^{S}(T), \qquad \lambda = 2 \left(\frac{1}{\epsilon} + i \right).
\eeqa
This solution reproduces Eq. (15) of \cite{Klauder} with the exponent $2/\epsilon$ there replaced by $ (2/\epsilon + i)$. (This correction is needed for the solution to satisfy Eq. (\ref{eqmo-kctcs-sap}) up to $\order(\epsilon^{0})$, although it does not affect the ensuring discussion.)

Klauder's stationary-action path is depicted in Fig. 1 in terms of $(p^{KS}(t),q^{KS}(t))$.

=== Fig. 1===

Remarkably the corresponding value of the action gives the desired exponent of (\ref{exact-cs-ta}).
\beqa
\frac{i}{\hbar} S_{KCS}^{SAP} := \frac{i}{\hbar} S_{KCS}[\bar{\xi}^{KS}, \xi^{KS}] = -\frac{1}{2} ( |\xi_{F}|^{2} + | \xi_{I} |^{2} ) + \xi^{*}_{F} \xi_{I} e^{-iT}+ \order(\epsilon).                \lab{sa-kctcs}    
\eeqa
Moreover it can be shown that 
\beqa
  S_{CS-\epsilon}[\bar{\xi}^{KS}, \xi^{KS}] = \order(\epsilon).
               \lab{sa-eterm}    
\eeqa
This follows from the property that the potentially dangerous integrand $\dot{\bar{\chi}}(t)\dot{\chi}(t)$ is proportional to $\epsilon^{-2}\exp(-2T/\epsilon)$, which is exponentially small. 
Hence, although Klauder's $\epsilon$-term plays an important role in 
ensuring the existence of, and determining, Klauder's stationary-action path, 
it does not contribute to the stationary value of the action.
In view of this circumstance, one might be tempted to follow what might be 
called the "semi-$\epsilon$ prescription"
\cite{fn4}:
\bqu
(i) Adopt Eq.~(\ref{eqmo-kctcs-sap}) and the boundary condition (\ref{bc-ctcs-qp2})
to determine $(\bar{\xi}^{KS}(t), \xi^{KS}(t))$.
\\
(ii) Once this has been done, discard $S_{CS-\epsilon}[\xi^{*}, \xi]$ altogether,
and work with $S_{CS}[\xi^{*}, \xi]$ alone to evaluate fluctuation integrals 
as well as the stationary value of the action.
\equ
Unfortunately, as shown in the next section, this "semi-$\epsilon$ prescription" fails;
integration over fluctuations leads to a non-sensical result.
The correct result 
\cite{fn5}
may be obtained only if Klauder's $\epsilon$-term is 
{\em properly discretized} and kept. 
There is no unique scheme of discretization, and the {\em proper discretization} amounts to working with the discrete-time formalism throughout
(see the end of Sec. V.B.).
 
\subsection{Klauder's Continuous-Time Spin-Coherent-State Path Integral}
In the case of the spin-coherent state, 
the relevant phase space is a sphere whose metric is $(d {\bf n})^2 (=(d\theta)^{2} + (\sin \theta)^{2}(d\phi)^{2}) $, or equivalently $4(1+|\xi|^{2})^{-2}|d\xi|^{2}$.
Klauder's augmented action, to be denoted by $S_{KSCS}[\xi^{*},\xi]$, now reads
\begin{mathletters}
 \beqa
 && S_{KSCS}[\xi^{*},\xi] 
       := S_{SCS-\epsilon}[\xi^{*},\xi] + S_{SCS}[\xi^{*},\xi],
               \lab{action-kctscs}\\    
 && S_{SCS-\epsilon}[\xi^{*},\xi] := \int_{0}^{T} dt \frac{i\hbar S \epsilon 
|\dot{\xi}(t)|^{2}}{(1+|\xi(t)|^{2})^{2}}. 
               \lab{action-eterm-kctscs} 
 \eeqa
 \label{klauder-scs-action}\end{mathletters}Klauder's stationary-action path specialized to the spin under the constant magnetic field obeys 
\begin{mathletters}
\beqa
&&\frac{\epsilon}{2}\left\{ \ddot{\bar{\xi}}~^{KS}(t) - \frac{2\xi^{KS}(t)(\dot{\bar{\xi}}~^{KS}(t))^{2}}{1 + \bar{\xi}^{KS}(t)\xi^{KS}(t) } \right\} + \dot{\bar{\xi}}~^{KS}(t) - i\bar{\xi}^{KS}(t) = 0 , \label{keqmo-scspi1}\\
&&\frac{\epsilon}{2}\left\{ \ddot{\xi}^{KS}(t) - \frac{2 \bar{\xi}^{KS}(t)(\dot{\xi}^{KS}(t))^{2}}{1 + \bar{\xi}^{KS}(t)\xi^{KS}(t) } \right\} - \dot{\xi}^{KS}(t) - i\xi^{KS}(t) = 0 . \label{keqmo-scspi2}
\eeqa
\label{keqmo-scspi}\end{mathletters}As in the previous section we employ the substitution (\ref{chixis1}),(\ref{chixis2}) to obtain the following equation which is correct up to $\order(\epsilon)$: 
\begin{mathletters}
\beqa
&& \frac{\epsilon}{2} \ddot{\bar{\chi}}(t) + (1+i\epsilon)\dot{\bar{\chi}}(t) - \frac{\epsilon R^{S}\chi(t)}{1+R^{S}\bar{\chi}(t)\chi(t)}(\dot{\bar{\chi}}(t) + 2i\bar{\chi}(t))\dot{\bar{\chi}}(t)  =0, \label{chieqmo-scspi1}\\
&& \frac{\epsilon}{2} \ddot{\chi}(t) - (1+i\epsilon)\chi(t) - \frac{\epsilon R^{S}\bar{\chi}(t)}{1+R^{S}\bar{\chi}(t)\chi(t)}(\dot{\chi}(t) - 2i\chi(t))\dot{\chi}(t)  = 0 \label{chieqmo-scspi2}
\eeqa
\label{chieqmo-scspi}\end{mathletters}where 
\beqa
R^{S} := \bar{\xi}^{S}(t)\xi^{S}(t) = \xi^{*}_{F}\xi_{I}e^{-iT}
\eeqa
is a constant.
The nonlinearity of the spin coherent state manifests itself in the last terms proportional to $R^{S}$.
Since the one in Eq. (\ref{chieqmo-scspi1}) is multiplied by $\dot{\bar{\chi}}(t)$, which vanishes except for the initial interval, we may replace $\chi(t)$ there by unity. (Recall that $\chi(t)$ should be essentially unity except for the final interval.) Similarly, $\bar{\chi}(t)$ in Eq. (\ref{chieqmo-scspi2}) may be replaced by unity.
We can straightforwardly solve the resulting equations to find 
\begin{mathletters}
\beqa
&&\bar{\chi}(t) = \frac{1 + R^{S}\bar{\chi}(0) + (\bar{\chi}(0) -1)e^{-\mu t}}{1 + R^{S} \bar{\chi}(0) -(\bar{\chi}(0)-1)R^{S}e^{-\mu t}} + \order(\epsilon), \label{solchi-scspi1} \\
&&\chi(t) = \frac{1+R^{S}\chi(T)+(\chi(T)-1)e^{-\mu (T-t)}}{1 + R^{S}\chi(T)-(\chi(T)-1)R^{S}e^{-\mu (T-t)}} +\order(\epsilon) , \label{solchi-scspi2}
\eeqa\label{solchi-scspi}\end{mathletters}where
\beqa
\bar{\chi}(0) = \xi^{*}_{I}/\bar{\xi}^{S}(0) ,\qquad \chi(T) = \xi_{F}/\xi^{S}(T),\qquad  \mu = 2\left( \frac{1}{\epsilon} + i \frac{1-R^{S}}{1+R^{S}} \right). \eeqa
This solution reproduces Eqs. (51-52) of \cite{Klauder} with the exponent $2/\epsilon$ there replaced by $\mu$.

The same mechanism as in the coherent-state case gives the result
\beqa
S_{SCS-\epsilon}[\bar{\xi}^{KS},\xi^{KS}] = \order(\epsilon).
\eeqa
Hence 
\beqa
S^{SAP}_{KSCS} &:=& S_{KSCS}[\bar{\xi}^{KS},\xi^{KS}] \nonumber \\
&=& i\hbar S \int_{0}^{T}dt \frac{R^{S}}{1+R^{S}\bar{\chi}(t)\chi(t)}(\bar{\chi}(t)\dot{\chi}(t)-\dot{\bar{\chi}}(t)\chi(t)) \nonumber \\
&+& \int_{0}^{T}dt \left\{i\hbar S \frac{\bar{\chi}(t)\chi(t)}{1+R^{S}\bar{\chi}(t)\chi(t)}(\bar{\xi}^{S}(t)\dot{\xi}^{S}(t) - \dot{\bar{\xi}}~^{S}(t)\xi^{S}(t))-{\cal H}(\bar{\xi}^{S}(t),\xi^{S}(t)) \right\}.
\eeqa
The first integral is essentially equal to 
\beqa
i\hbar S \int_{0}^{T}dt \left(\frac{R^{S}\dot{\chi}(t)}{1+R^{S}\chi(t)}-\frac{R^{S}\dot{\bar{\chi}}(t) }{1+R^{S}{\bar{\chi}}(t)} \right) 
= i\hbar S \ln \frac{(1+|\xi_{F}|^{2})(1+|\xi_{I}|^{2})}{(1+R^{S})^{2}},
\eeqa
while the second integral is the same as (\ref{specialcasevalue}) apart from a correction of $\order(\epsilon)$ since $\bar{\chi}(t)\chi(t)$ is essentially unity except for the initial and final intervals. The correct answer (\ref{exact-scs-ta}) is thus reproduced.
However one encounters a difficulty when it comes to evaluating the fluctuation integral (see the end of Sec. V.C.).

\section{Discrete-Time Formalism}
In order to resolve various difficulties encountered 
in the continuous-time formalism,
we go back to the basic definition of the amplitudes 
which the path integrals in question are supposed to represent.
Again it is instructive to begin by reviewing the familiar case
of the Feynman kernel.

\subsection{Discrete-Time Phase-Space Path Integral}
By a repeated use of the resolution of unity 
\beqa
\int dp |p\>\<p| = 1, \qquad \int dq |q\>\<q| = 1,
\eeqa
the Feynman kernel is expressed as
\begin{mathletters}
 \beqa
&& \< q_{F} | e^{ -i\hat{H}T/\hbar } | q_{I} \> 
   = \lim_{N \to \infty}
   \int \prod_{n=1}^{N-1} dq_{n} \prod_{n=1}^{N} \frac{dp_{n}}{2\pi\hbar} 
   \exp\left( \frac{i}{\hbar} {\cal S}_{PS}[ \{p\} , \{q\} ] \right) ,
   \lab{f-kernel-disps}\\
&& {\cal S}_{PS}[\{p\},\{q\}] := \sum_{n=1}^{N}
   \left\{( q_{n} - q_{n-1} )p_n - \varepsilon H ( p_{n} , q_{n} )  \right\}
   \lab{action-disps}
 \eeqa
 \label{discrete-cs-action}\end{mathletters}with
\beqa 
q_{N} \equiv q_{F}, \qquad q_{0} \equiv q_{I}, \qquad \varepsilon \equiv T/N,
\eeqa
and $\{p\}$ and $\{q\}$ standing for the set $\{p_1, p_2, \cdots, p_N \}$
and $\{q_1, q_2, \cdots, q_{N-1} \}$, respectively.
(To be precise ${\cal S}_{PS}$ should carry index $N$, 
which is omitted for brevity.) 
This is what we call the {\it discrete-time phase-space path integral} (DTPSPI). It is true that the time becomes effectively continuous 
in the limit $N \rightarrow \infty$.
But this should not blur the distinction from the continuous-time formalism.
What counts is that $N$ is kept finite until the very end of calculation.

For large $N$, one might be tempted to re-write the first term of (\ref{action-disps}) as
\beqa
\sum_{n=1}^{N} ( q_n - q_{n-1} ) p_n = 
\sum_{n=1}^{N} \varepsilon \frac{ q_n - q_{n-1} }{\varepsilon} p_n 
\sim \int_0^T dt \dot{q}(t)p(t),
\eeqa
thereby claiming to have reduced it to the first term of (\ref{ctps-action}).
But this argument is not warranted, 
because at this stage the integrand does not contain any factor  
which would ensure that $q_n - q_{n-1}$ is "small";
CTPSPI does {\em not} automatically follow from DTPSPI.

The multiple integral over the $2N -1$ variables may be evaluated 
by the stationary-phase method.
Let $( \{p^{cl}\}, \{q^{cl}\} )$ be the stationary point of the action 
${\cal S}_{PS}[\{p\}, \{q\}]$:
\begin{mathletters}
 \beqa
&& 0 = \left. \frac{\partial {\cal S}_{PS}[\{p\},\{q\}]}{\partial p_1} \right|_{cl} 
= q_1^{cl} - q_I - \left. \varepsilon \frac{\partial H(\{p\},\{q\})}{\partial p_1} \right|_{cl} , 
     \lab{}\\
&& 0 = \left. \frac{\partial {\cal S}_{PS}[\{p\},\{q\}]}{\partial p_n} \right|_{cl} 
=  q_n^{cl} - q_{n-1}^{cl} - \left. \varepsilon \frac{\partial H(\{p\},\{q\})}{\partial p_n} \right|_{cl}   : 2 \leq n \leq N-1,
     \lab{}\\
&& 0 = \left. \frac{\partial {\cal S}_{PS}[\{p\},\{q\}]}{\partial p_N} \right|_{cl} 
= q_F - q_{N-1}^{cl} - \left. \varepsilon \frac{\partial H(\{p\},\{q\})}{\partial p_N} \right|_{cl} , 
     \lab{}\\
&& 0 = \left. \frac{\partial {\cal S}_{PS}[\{p\},\{q\}]}{\partial q_n} \right|_{cl} 
= p_n^{cl} - p_{n+1}^{cl} - \left. \varepsilon \frac{\partial H(\{p\},\{q\})}{\partial q_n} \right|_{cl}  : 1 \leq n \leq N-1.
     \lab{}
 \eeqa
\label{}\end{mathletters}These constitute a set of $2N-1$ equations for the same number of unknowns.
There is no room for a "boundary condition" ; 
such a notion does not exist in the context of this set of equations.
On inspection one finds it convenient to {\em define} 
\beqa
    q_N^{cl} := q_F,  \qquad  q_0^{cl} := q_I.
                    \lab{bc-disps}
\eeqa  
(Remember that $q^{cl}_N$ and $q^{cl}_0$ did not exist among the variables.)
With this definition, the above set of equations takes the compact form:
\begin{mathletters}
 \beqa
&& q_n^{cl} - q_{n-1}^{cl} =   \left. ~~~\varepsilon \frac{\partial H(\{p\},\{q\})}{\partial p_n} \right|_{cl} \qquad : 1 \leq n \leq N,
          \lab{eqmo-disps-sap1}\\
&& p_{n+1}^{cl} - p_{n}^{cl} =   - \left. \varepsilon \frac{\partial H(\{p\},\{q\})}{\partial q_n} \right|_{cl}\qquad : 1 \leq n \leq N-1.
           \lab{eqmo-disps-sap2}
 \eeqa
\label{eqmo-disps-sap}\end{mathletters}One may regard this as a set of $2N-1$ equations 
for the $2N+1$ variables 
$\{p^{cl}_1, p^{cl}_2, \cdots, p^{cl}_N; q^{cl}_0, q^{cl}_1, \cdots, q^{cl}_N \}$,
and regard Eq.~(\ref{bc-disps}) as the boundary condition to be imposed.
In this way the notion of the boundary condition can be introduced, if desired,
merely as a matter of convenience \cite{fn1-1}.   
The factor $\varepsilon$ on the right-hand side ensures that 
$q_n^{cl} - q_{n-1}^{cl}$ and $p_{n+1}^{cl} - p_{n}^{cl}$ are 
of $\order (\varepsilon)$. 
Hence, in place of the above difference equation,
one may solve the differential equations (\ref{eqmo-ctps-sap}) with the boundary condition (\ref{bc-ctps2}).
The solution to the difference equation is then obtained as
\beqa
 q_n^{cl} =  q^{cl}(t_n) + \order (\varepsilon),
  \qquad  
 p_n^{cl} =  p^{cl}(t_n) + \order (\varepsilon),
           \lab{sol-sap-ctps}
\eeqa
where $t_n := n \varepsilon = (n/N)T$.
This is the rationale for Eqs.~(\ref{eqmo-ctps-sap}) and (\ref{bc-ctps2}) encountered in CTPSPI.

Although we mentioned in subsection III.A., that 
the evaluation of the fluctuation integral in CTPSPI is a routine matter,
it can be handled only after some "discretization", which however is not unique. By contrast, in DTPSPI, such a notion as "discretization" does not appear;
the fundamental formula is discrete by definition.
Hence fluctuation integral can be performed without ambiguity.
Decomposing the integration variables as
\beqa
  p_n =  p_n^{cl} + {\sl p}_n, \qquad q_n =  q_n^{cl} + {\sl q}_n,
           \lab{separated-path-disps}
\eeqa
one finds
\beq
{\cal S}_{PS}[\{p\},\{q\}] \simeq  S_{PS}^{cl} +  {\cal O}(\varepsilon)
                                 +   {\cal S}^{(2)}_{PS}[\{{\sl p}\},\{{\sl q}\}]
                       \lab{separated-action-disps}
\eeq
where $\simeq$ indicates that only terms up to the second order in 
fluctuations are kept. We have made use of  
\beqa
{\cal S}_{PS}[\{p^{cl}\},\{q^{cl}\}] = S_{PS}^{cl} +  
                             {\cal O}(\varepsilon)
                                  \lab{classical-action-disps}
\eeqa
with $S_{PS}^{cl} := S_{PS}[p^{cl},q^{cl}]$.
The Feynman kernel specialized to the harmonic oscillator now becomes
\beqa
\<q_F|e^{-i\hat{H}T/\hbar}|q_I\>&=& 
\exp\left( \frac{i}{\hbar}S_{PS}^{cl} \right)\lim_{N\to \infty} \int \prod_{n=1}^{N-1} d{\sl q}_{n} \prod_{n=1}^{N}
\frac{d{\sl p}_{n}}{2\pi\hbar}\nonumber \\ 
&&\times\exp \left[\frac{i}{\hbar}\sum_{n=1}^{N}\left\{ ({\sl q}_{n} - {\sl q}_{n-1}){\sl p}_{n} - \varepsilon \left(\frac{{\sl p}^{2}_{n}}{2} + \frac{{\sl q}^{2}_{n}}{2} \right) \right\} \right]
\eeqa
which involves Gaussian integrals.
However, each integration variable ${\sl p}_n$ appear in the exponent
in the form $i\varepsilon {\sl p}_n^2/\hbar$. 
Consequently it ranges effectively over the region
$|{\sl p}_n| \lsim  (\hbar/\varepsilon)^{1/2}$, which covers the entire real axis as 
$\varepsilon$ tends to 0 in the limit $N \rightarrow \infty$ .
Thus, ${\sl p}_n$'s cannot be regarded as constituting small fluctuation at all;
the picture that the phase-space path integral is dominated by
the classical path $(\{p^{cl}\},\{q^{cl}\})$ is erroneous. 
Fortunately, in the case of the harmonic oscillator (or of a non-relativistic particle in general),
integrations over ${\sl p}_n$'s can be carried out exactly.
The result is a configuration-space path integral,
in which ${\sl q}_n$'s can indeed be regarded as constituting small fluctuation 
since they appear in the exponent in the form $i{\sl q}_n^2/(\hbar\varepsilon)$.

\subsection{Discrete-Time Coherent-State Path Integral}
In dealing with the case of the coherent-state path integral,
it is more convenient to work with $\xi$ related to $(p,q)$ via Eq.~(\ref{xi-q-p}).
By a repeated use of the resolution of unity 
\beq
\int \frac{d\xi d\xi^{*}}{2\pi i}|\xi\>\<\xi| = 1  ,\qquad \frac{d\xi d\xi^{*}}{2\pi i} := \frac{d(\Re \xi)d(\Im \xi)}{\pi} = \frac{dpdq}{2\pi\hbar},
\eeq
the amplitude in question is expressed as
\begin{mathletters}
\beqa
&& \< \xi_{F} | e^{-i\hat{H}T/\hbar} | \xi_{I} \> 
= \lim_{N \to \infty}\int \prod_{n=1}^{N-1} \frac{d\xi_{n}d\xi^{*}_{n}}{2 \pi i} \exp\left( \frac{i}{\hbar}{\cal S}_{CS}[  \{\xi^{*}\} ,\{ \xi \}  ] \right) ,
             \label{ta-discs} \\
&& \frac{i}{\hbar}{\cal S}_{CS}[  \{\xi^{*}\} ,\{ \xi \}  ] = \sum_{n=1}^{N} \left[
 -\frac{1}{2}( |\xi_{n}|^{2} + |\xi_{n-1}|^{2} ) + \xi^{*}_{n} \xi_{n-1} -\frac{i}{\hbar}\varepsilon {\cal H}( \xi^{*}_{n} , \xi_{n-1} )\right]  ,
             \label{action-discs}
 \eeqa
\end{mathletters}where the convention in the previous subsection is followed, that is,
\beq
\xi_{N} \equiv \xi_{F} ,\qquad \xi_{0} \equiv \xi_{I}, \qquad \varepsilon \equiv N/T, 
\eeq
and $\{\xi\}$ stands for the set $\{\xi_1, \xi_2, \cdots, \xi_{N-1} \}$
of $N-1$ complex variables.
Since each $\xi_n^*$ is the complex conjugate of $\xi_n$,
the notation ${\cal S}_{CS}[\{ \xi^* \}, \{ \xi \}]$ is redundant at this stage, but will be found useful later. 

\subsubsection{stationary-action path}
Let $( \{\bar{\xi}^S \}, \{\xi^S \} )$ be the stationary point of the action 
${\cal S}_{CS}[\{\xi^* \}, \{\xi \}]$:
\begin{mathletters}
 \beqa
&& 0=  \left. \frac{i}{\hbar}\frac{\delta {\cal S}_{CS} }{\delta \xi^{*}_{n} } \right|_{S} = \left. \left( - \xi_{n} + \xi_{n-1} - \frac{i}{\hbar}\varepsilon\frac{\partial {\cal H}}{\partial \xi^{*}_{n}} \right)\right|_{S}  \qquad :2 \le n \le N-1. \label{sabun-discs1}\\
&& 0= \left.\frac{i}{\hbar} \frac{\delta {\cal S}_{CS} }{\delta \xi^{*}_{1} } \right|_{S} = \left. \left( - \xi_{1} + \xi_{I} - \frac{i}{\hbar}\varepsilon\frac{\partial {\cal H}}{\partial \xi^{*}_{1}} \right)\right|_{S}  . 
   \label{sabun-discs2}\\
&& 0= \left. \frac{i}{\hbar}\frac{\delta {\cal S}_{CS} }{\delta \xi_{n} } \right|_{S} = \left. \left( - \xi^{*}_{n} + \xi^{*}_{n+1} - \frac{i}{\hbar}\varepsilon\frac{\partial {\cal H}}{\partial \xi_{n}} \right)\right|_{S} \qquad :1 \le n \le N-2 .    \label{sabun-discs3}\\
&& 0= \left. \frac{i}{\hbar}\frac{\delta {\cal S}_{CS} }{\delta \xi_{N-1} } \right|_{S} = \left. \left( - \xi^{*}_{N-1} + \xi^{*}_{F} - \frac{i}{\hbar}\varepsilon\frac{\partial {\cal H}}{\partial \xi_{N-1}} \right)\right|_{S}  .  
   \label{sabun-discs4}
\eeqa
\label{sabun-discs10}\end{mathletters}These constitute a set of $2(N-1)$ equations for the same number of unknowns.
Again, there is no room for a "boundary condition".
On inspection one finds it convenient to {\em define} 
\beq
    \qquad \bar{\xi}^{S}_{N} := \xi^{*}_{F}. \qquad  \xi^{S}_{0} := \xi_{I}.   \lab{bc-discs}
\eeq  
(Remember that $\bar{\xi}_N^S$ and $\xi_0^S$ did not exist 
among the unknowns.)
With this definition, the above set of equations takes the compact form:
\begin{mathletters}
\beqa
\xi^{S}_{n} - \xi^{S}_{n-1} &=& -\frac{i}{\hbar}\varepsilon  \left.\frac{\partial {\cal H}}{\partial \xi^{*}_{n}} \right|_{S}  \qquad :1 \le n \le N-1 ,  
\label{sabun-discs5}\\
\bar{ \xi } ^{S}_{n+1} - \bar{ \xi } ^{S}_{n} &=& ~~~ \frac{i}{\hbar}\varepsilon  \left.\frac{\partial {\cal H}}{\partial \xi_{n}} \right|_{S}  \qquad :1 \le n \le N-1 .  
\label{sabun-discs6}
\eeqa
\label{sabun-discs20}\end{mathletters}One may regard this as a set of $2N-2$ equations 
for the $2N$ unknowns 
$\{ \bar{\xi}_1^S, \bar{\xi}_2^S, \cdots, \bar{\xi}_{N}^S ; 
\xi_0^S, \xi_1^S, \cdots, \xi_{N-1}^S \}$,
and regard Eq.~(\ref{bc-discs}) as the boundary condition to be imposed.
The factor $\varepsilon$ on the right-hand side ensures that 
$\xi_{n}^S - \xi_{n-1}^S$ and $\bar{\xi}_{n+1}^S - \bar{\xi}_{n}^S$ are 
of $\order (\varepsilon)$. 
Hence, in place of the above difference equation,
one may solve the differential equations (\ref{eqmo-ho-ctcs-xi}) 
with the boundary condition (\ref{bc-ctcs-xi}).
The solution to the difference equation is then obtained as
\begin{mathletters}
\beqa
\xi^{S}_{n} &=& \xi^{S}(t_{n}) + \order(\varepsilon) \qquad :0 \le n \le N-1 ,
\lab{sol-diffeq-cs1}\\
\bar{\xi}^{S}_{n} &=& \bar{\xi}^{S}( t_{n} ) + \order(\varepsilon) \qquad :1 \le n \le N. \lab{sol-diffeq-cs2}
\eeqa
\label{sol-diffeq-cs}\end{mathletters}This is the rationale for Eqs.~(\ref{eqmo-ho-ctcs-xi}) and (\ref{bc-ctcs-xi}) encountered in CTCSPI.

In what follows the stationary point shall be called 
the stationary-action path.
However, it is important to note that $\xi_F$ and $\xi_I^*$ do not occur 
in Eqs.~(\ref{sabun-discs20}) and that these equations do not involve 
such unknowns as $\xi_{N}^S$ or $\bar{\xi}_0^S $.
Hence neither $\xi_F - \xi_{N-1}^S$ nor $\bar{\xi}_1^S - \xi_I^*$ 
can be said to be of $\order(\varepsilon)$.
Consequently, even in the limit $N \rightarrow \infty$,
 the stationary-action path need not be continuous 
at the initial and  final times.

Let us illustrate this point with the harmonic oscillator, for which
 Eqs.~(\ref{sabun-discs20}) reduce to
\beqa
\xi^{S}_{n} - \xi^{S}_{n-1} = -i\varepsilon  \xi^{S}_{n-1} ,\qquad
\bar{ \xi } ^{S}_{n+1} - \bar{ \xi } ^{S}_{n} = i\varepsilon  \bar{\xi}^{S}_{n+1} \qquad :1 \le n \le N-1,
\eeqa
and the solution is immediately found  as 
\beqa
\xi^{S}_{n} = (1 - i\varepsilon)^{n} \xi_{I}  = \xi_{I} e^{-it_{n}} + \order(\varepsilon), \qquad \bar{\xi}^{S}_{n} = ( 1-i\varepsilon)^{N-n}\xi^{*}_{F} = \xi^{*}_{F}e^{ i(t_{n} - T )}+ \order(\varepsilon) . \lab{sap-discs}
\eeqa
The last approximate expression agrees with the result obtained via the procedure (\ref{sol-diffeq-cs}) of course.
In terms of $ ( \{p^S\}, \{q^S\} )$ related to 
$( \{\xi^S\},  \{ {\bar \xi}^S \} )$ via (\ref{sap-xi-q-p}), 
the stationary-action path is given by
\begin{mathletters}
\beqa
q^{S}_{n} &=& ~~~~\sqrt{\frac{\hbar}{2}} ( \xi^{S}_{n} + \bar{\xi}^{S}_{n} ) 
=~~~
\sqrt{\frac{\hbar}{2}}  ( \xi_{I} e^{-i t_{n} } + \xi^{*}_{F} e^{ i  ( t_{n} - T) } )\qquad    :1 \le n \le N-1,\\
p^{S}_{n} &=& -i\sqrt{\frac{\hbar }{2}} ( \xi^{S}_{n} - \bar{\xi}^{S}_{n} )  
=
-i\sqrt{\frac{\hbar }{2}}( \xi_{I} e^{-i t_{n} } - \xi^{*}_{F} e^{ i  ( t_{n} - T) } )\qquad  :1 \le n \le N-1.
\eeqa
\end{mathletters}(Note that neither $ ( p_N^S, q_N^S )$ nor $ ( p_0^S, q_0^S )$ is defined.)
This is depicted in Fig.~2. 

=== Fig. 2 ===

The stationary-action path is not in general real as observed in Sec.~III.B..
Its real and imaginary parts trace a circle of radius 
\beqa
\left[\frac{\hbar}{2}\{|\xi_{F}|^{2}+|\xi_{I}|^{2} \pm 2 \Re (\xi^{*}_{F}\xi_{I}e^{-iT})\}\right]^{1/2},
\eeqa
respectively. Neither of them coincides with the classical path 
connecting $(p_I, q_I)$ and $(p_F, q_F)$ in time $T$; 
such a classical path would exist only in the special case (\ref{special-bc}) alone. 

Comparison of Figs.~1 and 2 shows 
that Klauder's $\epsilon$-prescription is a device to interpolate 
smoothly between $(p_I, q_I)$ and $ ( p_1^S, q_1^S )$ 
as well as between $(p_{N-1}^S, q_{N-1}^S)$ and $ ( p_F, q_F )$.

\subsubsection{$\varepsilon$-term of the action}
The action can be re-written as
\begin{mathletters}
 \beqa
&& {\cal S}_{CS}[\{\xi^* \}, \{\xi \}] = 
        {\cal S}_{CS-\varepsilon}[\{\xi^* \}, \{\xi \}] 
           + {\cal S}_{CS-c}[\{\xi^* \}, \{\xi \}]
                +{\cal S}_{CS-d}[\{\xi^* \}, \{\xi \}] ,
                                    \label{separated-action-discs} \\
&& \frac{i}{\hbar}{\cal S}_{CS-\varepsilon}[\{\xi^* \}, \{\xi \}] := -\frac{1}{2}\left(
|\xi_{F}-\xi_{N-1}|^{2} + \sum_{n=2}^{N-1}|\xi_{n}-\xi_{n-1}|^{2}+|\xi_{1}-\xi_{I}|^{2} \right), \lab{eterm-action-discs} \\
&& \frac{i}{\hbar}{\cal S}_{CS-c}[\{\xi^* \}, \{\xi \}] := -\frac{1}{2}\bigg( \xi^{*}_{F}(\xi_{F}-\xi_{N-1})-(\xi^{*}_{F}-\xi^{*}_{N-1})\xi_{F} \nonumber \\
&& + \sum_{n=2}^{N-1}\{  \xi^{*}_{n}(\xi_{n}-\xi_{n-1})-(\xi^{*}_{n}-\xi^{*}_{n-1})\xi_{n}\} + \xi^{*}_{1}(\xi_{1}-\xi_{I})-(\xi^{*}_{1} - \xi^{*}_{I})\xi_{I} \bigg), \lab{cterm-action-discs} \\
&& {\cal S}_{CS-d}[\{\xi^* \}, \{\xi \}]:= -\varepsilon \sum_{n=1}^{N}{\cal H}(\xi^{*}_{n},\xi_{n-1}).    \lab{dterm-action-discs}
  \eeqa
\end{mathletters}The first term would resemble Klauder's $\epsilon$-term
if the following manipulation were correct.
\beqa
   {\cal S}_{CS-\varepsilon}[\{\xi^* \}, \{\xi \}] 
    = \frac{i\hbar}{2}\sum_{n=1}^{N}\varepsilon^{2} \frac{|\xi_{n} - \xi_{n-1}|^{2}}{\varepsilon^{2}} \sim \frac{i\hbar}{2} \varepsilon \int_{0}^{T} dt \dot{\xi}^{*}(t)\dot{\xi}(t).                    \lab{keterm-action}
\eeqa
It is of course strange to keep $\varepsilon$ partially
while converting a sum into an integral
under the supposition that everything behaves smoothly
in the limit $\varepsilon \rightarrow 0$. 
Nevertheless, this is the only clue to identify what would correspond
to Klauder's $\epsilon$-term.
Therefore we call ${\cal S}_{CS-\varepsilon}[\{\xi^* \}, \{\xi \}]$
the $\varepsilon$-term, which is the reason for the notation adopted. 
Likewise the second term would resemble the first term of (\ref{action-ctcs-xi}). Hence it is  to be called the canonical term. The last term represents the contribution of the Hamiltonian and is to be called the dynamical term.

\subsubsection{stationary action}
We now evaluate the contributions of the three terms of (\ref{separated-action-discs}) separately
to the stationary action.

The contribution of the $\varepsilon$-term may be further 
separated into three parts.
That from the intermediate times (the second term of (\ref{eterm-action-discs})) is obviouly
of $\order(\varepsilon)$ in view of (\ref{sol-diffeq-cs}):
\begin{mathletters}
\beqa
\sum_{n=2}^{N-1}(\bar{\xi}^{S}_{n}-\bar{\xi}^{S}_{n-1})(\xi^{S}_{n}-\xi^{S}_{n-1}) = \varepsilon \int_{0}^{T} dt \dot{\bar{\xi^{S}}}(t)\dot{\xi}^{S}(t) = \order(\varepsilon). 
\eeqa
The "initial discontinuity" (the last term of (\ref{eterm-action-discs})) contributes
\beqa
   (\bar{\xi}^{S}_{1}-\xi^{*}_{I})(\xi^{S}_{1} - \xi_{I}) = 
(\xi^{*}_{F}e^{i(\varepsilon - T)}-\xi^{*}_{I})(e^{-i\varepsilon}-1)\xi_{I} = \order(\varepsilon).
\eeqa
\end{mathletters}Similar result is found for the contribution of 
the "final discontinuity" (the first term of (\ref{eterm-action-discs})).
Hence
\beqa
      {\cal S}_{CS-\varepsilon}^{SAP}
    := {\cal S}_{CS-\varepsilon}[\{ \bar{\xi}^S \}, \{ \xi^S \}]
      = \order(\varepsilon).
\eeqa
It is thus concluded that the $\varepsilon$-term does not contribute 
to the stationary action,
which is in accord with the result of Klauder that his $\epsilon$-term does not contribute to the stationary action.
 One should not be betrayed by the expression (\ref{eterm-action-discs}), from which one might guess that the discontinuities shown in Fig.~2 would make a contribution of $\order(\varepsilon^{0})$. The reason why they do not is that $\bar{\xi}^{S}_{n}$ and $\xi^{S}_{n}$ are not necessarily mutually complex conjugate.
 
The contribution of the canonical term may be evaluated in a similar fashion. Thus, the intermediate times contribute
\begin{mathletters}\beqa
-\frac{1}{2}\sum_{n=2}^{N-1}\{\bar{\xi}^{S}_{n}(\xi^{S}_{n}-\xi^{S}_{n-1})-(\bar{\xi}^{S}_{n} -\bar{\xi}^{S}_{n-1})\xi^{S}_{n} \} = iT \xi^{*}_{F}\xi_{I}e^{-iT}+\order(\varepsilon),\label{canonical-inter-discs}
\eeqa
while the final and initial discontinuity respectively contributes
\beqa
&& -\frac{1}{2}\{ \xi^{*}_{F}(\xi_{F}- \xi^{S}_{N-1})
-(\xi^{*}_{F}-\bar{\xi}^{S}_{N-1})\xi_{F}\} 
= \frac{1}{2}\xi^{*}_{F}\xi_{I}e^{-iT} - \frac{1}{2}|\xi_{F}|^{2} + \order(\varepsilon),  \lab{final-tran-discs} \\
&& -\frac{1}{2}
\{ \bar{\xi}^{S}_{1} (\xi^{S}_{1} - \xi_{I})
-(\bar{\xi}^{S}_{1} - \xi^{*}_{I} ) \xi^{S}_{1} \} 
= \frac{1}{2} \xi^{*}_{F} \xi_{I} e^{-iT} - \frac{1}{2}|\xi_{I}|^{2} + \order(\varepsilon). \lab{initial-tran-discs} 
\eeqa
\end{mathletters}Finally, the contribution of the dynamical term is found as
\beqa
-i\varepsilon\sum_{n=1}^{N}\bar{\xi}^{S}_{n}\xi^{S}_{n-1} = -iT\xi^{*}_{F}\xi_{I}e^{-iT} + \order(\varepsilon),
\eeqa
which cancels the intermediate-time contribution (\ref{canonical-inter-discs}) of the canonical term. Hence the stationary action , to be denoted by, ${\cal S}^{SAP}_{CS}$, is determined entirely by the discontinuity parts of the canonical term as 
\beqa
\frac{i}{\hbar}{\cal S}^{SAP}_{CS} &:=& \frac{i}{\hbar}{\cal S}_{CS}[\{ \bar{\xi}^{S} \},\{ \xi^{S} \}] \\
                                    &=& -\frac{1}{2}(|\xi_{F}|^{2}+|\xi_{I}|^{2}) + \xi_{F}^{*} \xi_{I} e^{-iT} + \order(\varepsilon),
\eeqa 
which reproduces the exponent of (\ref{exact-cs-ta}).

\bigskip
Comment: One might start from the following action \cite{Itzykson}
\beqa
&&\frac{i\hbar}{2}\{ \xi^{*}_{F}(\xi_{F}-\xi_{N-1})-(\xi^{*}_{1}-\xi^{*}_{I})\xi_{I}\}\nonumber \\
 &&+ \int_{0}^{T}dt\left\{\frac{i\hbar}{2}( \xi^{*}(t)\dot{\xi}(t) - \dot{\xi}^{*}(t)\xi(t) ) - {\cal H}(\xi^{*}(t),\xi(t)) \right\},
\eeqa
which is a mixture of discrete and continuous forms. 
Adopting (\ref{sol-ho-ctcs}) as the stationary-action path, 
one would obtain the correct value for the stationary action.
However, re-writing the action (\ref{separated-action-discs}) in such a mixed form is not a unique procedure.
Success would not be guaranteed unless one knew the answer beforehand.
At any rate one would fail if one proceeded to integration over fluctuations.

\subsubsection{fluctuations}
Let us turn to the evaluation of fluctuation integrals,
again keeping track of the roles played by each of the terms of (\ref{separated-action-discs}).
Separating the integration variables as
\beqa
\xi_{n} = \xi^{S}_{n} + \eta_{n} ,\qquad \xi^{*}_{n} = \bar{\xi}^{S}_{n} + \eta^{*}_{n} \qquad  :1 \le n \le N-1 , \lab{decomposed-xi-disscs}
\eeqa
one finds
\begin{mathletters}\beqa
 &&  {\cal S}_{CS}[\{\xi^* \}, \{\xi \}] = 
        {\cal S}_{CS}^{SAP} + {\cal O}(\varepsilon)
         +  {\cal S}^{(2)}_{CS}[\{\eta^* \}, \{\eta \}],
                 \lab{separared-sap-action-discs}\\
 && {\cal S}^{(2)}_{CS}[\{\eta^* \}, \{\eta \}] :=
          {\cal S}^{(2)}_{CS-\varepsilon}[\{\eta^* \}, \{\eta \}] 
           + {\cal S}^{(2)}_{CS-cd}[\{\eta^* \}, \{\eta \}],
                       \lab{separated-fluc-action-discs}
\eeqa
\end{mathletters}where
\begin{mathletters}\beqa
  && \frac{i}{\hbar}{\cal S}^{(2)}_{CS-\varepsilon}[\{\eta^* \}, \{\eta \}] 
:=  -\sum_{n=1}^{N-1} \eta^{*}_{n}\eta_{n} 
   + \frac{1}{2}\sum_{n=2}^{N-1}( \eta^{*}_{n}\eta_{n-1} + \eta^{*}_{n-1}\eta_{n} ) ,
                         \lab{eterm-fluc-action-discs}\\
  && \frac{i}{\hbar}{\cal S}^{(2)}_{CS-cd}[\{\eta^* \}, \{\eta \}] 
:= \frac{1}{2}\sum_{n=2}^{N-1}( \eta^{*}_{n}\eta_{n-1} - \eta^{*}_{n-1}\eta_{n} ) - i\varepsilon \sum_{n=2}^{N-1}\eta^{*}_{n}\eta_{n-1}.                     
\lab{usual-term-fluc-action-discs}
\eeqa
\end{mathletters}Accordingly 
\begin{mathletters}
\beqa
&& \<\xi_{F}|e^{-i\hat{H}T/\hbar}|\xi_{I}\> = 
\exp\left(\frac{i}{\hbar} {\cal S}_{CS}^{SAP} \right){\cal K}^{(2)}_{CS}(T), \\
&& {\cal K}^{(2)}_{CS}(T) := \lim_{N \to \infty} 
                 \int \prod_{n=1}^{N-1} 
                   \frac{d \eta_{n} d \eta^{*}_{n} }{2\pi i} 
                     \exp \left( \frac{i}{\hbar}{\cal S}^{(2)}_{CS}[\{\eta\},\{\eta^{*}\}] \right).   
\eeqa
\end{mathletters}At this stage $\eta^*$ is the complex conjugate of $\eta$, and 
\beqa
\frac{d\eta_{n}d\eta^{*}_{n}}{2\pi i} := \frac{d{\sl p}_{n}d{\sl q}_{n}}{2\pi\hbar},
\eeqa
with $({\sl p}_{n},{\sl q}_{n})$ related to $(\eta^{*}_{n},\eta_{n})$ via Eq.~(\ref{xi-q-p}).

Noting that
\beqa
   \frac{i}{\hbar}{\cal S}^{(2)}_{CS}[\{\eta^* \}, \{\eta \}] 
= - \sum_{n=1}^{N-1}  \eta^{*}_{n} \left\{ \eta_{n} - \alpha \eta_{n-1} \right\}, \qquad \alpha := 1 -i\varepsilon,
\eeqa
where we have defined $\eta_0 = 0$ 
\cite{fn6},
we make further change of the integration variables as \cite{1}
\beq
\eta^{'*}_{n} := \eta^{*}_{n} ,\qquad  \eta^{'}_{n} := \eta_{n} - \alpha \eta_{n-1} \label{variable-change}
\eeq
to find
\beqa
{\cal K}^{(2)}_{CS}(T) = \lim_{N \to \infty} \int \prod_{n=1}^{N-1} \frac{d \eta^{'}_{n} d \eta^{'*}_{n} }{2\pi i} \exp\left( -\sum_{n=1}^{N-1}\eta^{'*}_{n}\eta^{'}_{n}\right) = 1. \label{flucint-disscspi}
\eeqa
The complete amplitude (\ref{exact-cs-ta}) is thus recovered by treating DTCSPI in the stationary-action approximation, which should be exact for the harmonic oscillator. 
Note that $\eta_{n}$ appears in the exponent of the integral in the form $-|\eta_{n}|^{2} ( = ( {\sl p}_{n}^{2} + {\sl q}_{n}^{2} )/2\hbar )$. Hence the effective range of integration over ${\sl p}_{n}$ and ${\sl q}_{n}$ is of $\order(\hbar^{1/2})$. Thus, contrary to the case of DTPSPI, $\eta_{n}$'s constitute a small fluctuation in the quasi-classical situation.

What would happen if the $\varepsilon$-term were discarded 
in integrating over the fluctuations with the spirit of the "semi-$\epsilon$-prescription mentioned at Sect. IV.A.?
The integral (\ref{flucint-disscspi}) would then be replaced by
\beqa
{\cal K}^{(2)}_{CS-cd}(T) := 
\int \prod _{n=1}^{N-1} \frac{d\eta_{n}d\eta^{*}_{n}}{2\pi i} 
\exp \left( \frac{i}{\hbar}{\cal S}^{(2)}_{CS-cd}[\{\eta^{*}\},\{\eta\}]\right).
\eeqa
In order to evaluate this integral, we may write ${\cal S}^{(2)}_{CS-cd}$ in the following matrix expression:
\beqa
\frac{i}{\hbar}{\cal S}^{(2)}_{CS-cd} = -\frac{1}{2}~ ^{t}{\bf \eta}^{*} {\cal M} {\bf \eta}
\eeqa
where
\beqa
&& ^{t}{\bf \eta}^{*} := 
\left(
	\begin{array}{cccc}
\eta^{*}_{1} & \eta^{*}_{2} & \ldots & \eta^{*}_{N-1}
\end{array}\right), \qquad
{\cal M} :=
\left(
	\begin{array}{cccc}
0  & 1      &        &         \\
-a & \ddots & \ddots &         \\
   & \ddots & \ddots &   1     \\
   &        &  -a    &   0    
\end{array}\right),
\qquad
{\bf \eta} :=
\left(
	\begin{array}{c}
\eta_{1} \\
\eta_{2} \\
\vdots   \\
\eta_{N-1}
\end{array}\right), \label{matrix-exp}
\eeqa
with
 $ a := 1 -2i\varepsilon \simeq e^{-2i\varepsilon}$. The determinant of ${\cal M}$ is given as 
\beqa
\det {\cal M} =
\left\{
	\begin{array}{cc}
a^{\frac{N-1}{2}} & :{\rm if}~  N ~{\rm is ~odd} \\
0                 & ~:{\rm if}~  N ~{\rm is ~even} 
\end{array}\right. .
\eeqa
Hence, $N$ must be chosen to be an odd integer in order for the integral to make sence. With this coice, 
\beqa
{\cal K}^{(2)}_{CS-cd}(T) &=& \lim_{N \to \infty} 2^{N-1}( \det {\cal M} )^{-1} \nonumber \\
                          &=& e^{-iT}\lim_{N \to \infty}2^{N-1},
\eeqa
which does not even tend to a finite value in the limit $N \rightarrow \infty$.
 This is the non-sensical result announced in the introduction.

If one started from Klauder's action (\ref{action-kctscs}) and make the formal 
expansion as Eq. (\ref{expand-xiseta}), then one could find Eq. (\ref{expand-sapfluc}) with the subscript $CS$ replaced by $KCS$, where
\beqa
S^{(2)}_{KCS}[\eta^{*},\eta] := -i\hbar \int_{0}^{T}dt \left\{-\frac{1}{2}\epsilon \dot{\eta}^{*}(t)\dot{\eta}(t)-\frac{1}{2}(\eta^{*}(t)\dot{\eta}(t)-\dot{\eta}^{*}(t)\eta(t))-i\eta^{*}(t)\eta(t)\right\}.
\eeqa
One might adopt the following discretization 
\beqa
S^{(2)}_{KCS}[\eta^{*},\eta] &\to& -i\hbar \sum_{n=1}^{N}
\left[
-\frac{1}{2} 
(\eta^{*}_{n}-\eta^{*}_{n-1})(\eta_{n}-\eta_{n-1}) \right. \nonumber \\
&&-\left.\frac{1}{2} \left\{\eta^{*}_{n}(\eta_{n}-\eta_{n-1}) -(\eta^{*}_{n}-\eta^{*}_{n-1}) \eta_{n} \right\} -i\varepsilon\eta^{*}_{n}\eta_{n-1}\right], \\
 {\cal D}\eta {\cal D}\eta^{*} &\to& \prod_{n=1}^{N-1}\frac{d\eta_{n}d\eta^{*}_{n}}{2\pi i}.
\label{discretization}
\eeqa
The fluctuation integral would then give unity. Clealy, adoption of the above discretization scheme is equivalent to working with DTCSPI.
 However there is no compelling reason why we should adopt the particular discretization. If we adopted ( with $\eta^{*}_{N+1}= \eta_{-1} = 0$, )
\beqa
S^{(2)}_{KCS}[\eta^{*},\eta] &\to& 
-i\hbar\sum_{n=1}^{N}\bigg[
-\frac{1}{2} 
(\eta^{*}_{n+1}-\eta^{*}_{n})(\eta_{n}-\eta_{n-1}) \nonumber \\
&&- \frac{1}{2} \left\{\eta^{*}_{n}(\eta_{n}-\eta_{n-1})
-(\eta^{*}_{n}-\eta^{*}_{n-1}) \eta_{n-1} \right\} -i\varepsilon\eta^{*}_{n}\eta_{n-1}\bigg] \nonumber \\
 &=& \frac{i\hbar}{2}\sum_{n=1}^{N-1}\eta_{n}^{*}( \eta_{n} + 2i\varepsilon\eta_{n-1} - \eta_{n-2} ) .
\eeqa
we would obtain a non-sensical result:
\beqa
\int {\cal D}\eta{\cal D}\eta^{*} \exp\left(\frac{i}{\hbar}S^{(2)}_{KCS}[\eta^{*},\eta] \right) \to  \lim_{N\to\infty}2^{N-1}.
\eeqa
We conclude that the fluctuation integral can be unambiguously evaluated only in DTCSPI.

\subsection{Discrete-Time Spin-Coherent-State Path Integral}
The case of spin-coherent-state path integral may be discussed 
in parallel with the coherent-state case.
Thus, we work with $\xi$ related to $\theta$ and $\phi$ via Eq.~(\ref{xi-phi-theta}).
By a repeated use of the resolution of unity 
\beq
\int \frac{2S+1}{2\pi i}\frac{d \xi d\xi^{*}}{( 1 + |\xi|^{2} )^{2}}|\xi\>\<\xi|  = 1 ,
\eeq
the amplitude in question is expressed as
\begin{mathletters}\beqa
&&\<\xi_{F}|e^{-i\hat{H}T/\hbar}|\xi_{I}\> 
= \lim_{N \to \infty}\int \prod_{n=1}^{N-1} \frac{2S + 1}{2\pi i} \frac{d\xi_{n}d\xi^{*}_{n}}{ ( 1 + |\xi_{n}|^{2})^{2} } 
\exp\left( \frac{i}{\hbar}{\cal S}_{SCS}[  \{\xi^{*}\} ,\{ \xi\}  ] \right),  
     \lab{ta-disscs}\\
&& \frac{i}{\hbar}{\cal S}_{SCS}[  \{\xi^{*}\} ,\{ \xi \}  ] = \sum_{n=1}^{N} \left[ S \ln \frac{ ( 1 + \xi^{*}_{n}\xi_{n-1} )^{2} }{ ( 1+ |\xi_{n}|^{2} ) ( 1+ |\xi_{n-1}|^{2} ) } - \frac{i}{\hbar}\varepsilon {\cal H}( \xi^{*}_{n} , \xi_{n-1} )  \right],\lab{action-disscs}
\eeqa
\end{mathletters}where the convention in the previous subsection is followed.
The remark to the notation ${\cal S}_{CS}[\{ \xi^* \}, \{ \xi \}]$ 
applies here as well. 

\subsubsection{stationary-action path}
As in the case of DTCSPI, 
we denote the stationary-action path (i.e., the stationary point of the action)
by $( \{\xi^S \}, \{\bar{\xi}^S\} )$. It obeys a set of equations whose 
 basic structure is the same as that of Eqs.~(\ref{sabun-discs10}).
In particular, there is no room for a "boundary condition" .
Adopting the {\em definition} (\ref{bc-discs}),
We can cast the set of equations into the following form:
\begin{mathletters}\beqa
&&2S \frac{\xi^{S}_{n} - \xi^{S}_{n-1}}{(1+\bar{\xi}^{S}_{n}\xi^{S}_{n-1})(1+\bar{\xi}^{S}_{n}\xi^{S}_{n})} 
~=~ - \left.\frac{i}{\hbar}\varepsilon\frac{\partial {\cal H}}{\partial \xi_{n}} \right|_{S}  \qquad :1 \le n \le N-1,
          \lab{sabun-disscs5}\\
&&2S \frac{\bar{\xi}^{S}_{n+1} - \bar{\xi}^{S}_{n}}{(1+\bar{\xi}^{S}_{n+1}\xi^{S}_{n})(1+\bar{\xi}^{S}_{n+1}\xi^{S}_{n+1})} =  \left.\frac{i}{\hbar}\varepsilon\frac{\partial {\cal H}}{\partial \xi^{*}_{n}} \right|_{S}
  \qquad :1 \le n \le N-1.
           \lab{sabun-disscs6}
\eeqa
\end{mathletters}Again, the factor $\varepsilon$ on the right-hand side ensures that 
$\xi_{n}^S - \xi_{n-1}^S$ and $\bar{\xi}_{n+1}^S - \bar{\xi}_{n}^S$ are 
of $\order (\varepsilon)$,
but neither $\xi_F - \xi_{N-1}^S$ nor $\bar{\xi}_1^S - \xi_I^*$ 
can be said to be of $\order(\varepsilon)$.

For the spin under a constant magnetic field
described by the Hamiltonian (\ref{scs-constmag-hamiltonian}),
the above equations reduce to
\begin{mathletters}\beqa
&& \frac{ \xi^{S}_{n} - \xi^{S}_{n-1} }
{ 1 + \bar{\xi}^{S}_{n}\xi^{S}_{n} } 
=  
~-i\varepsilon  
\frac{ \xi^{S}_{n-1} }
{  1 + \bar{\xi}^{S}_{n} \xi^{S}_{n-1} } 
\qquad :1 \le n \le N-1,  \label{sabun-disscs7} \\
&& \frac{ \bar{\xi}^{S}_{n} - \bar{\xi}^{S}_{n-1} }
{ 1 + \bar{\xi}^{S}_{n-1} \xi^{S}_{n-1} } 
=  i\varepsilon  
\frac{ \bar{\xi}^{S}_{n} }
{  1 + \bar{\xi}^{S}_{n} \xi^{S}_{n-1} } 
\qquad :2 \le n \le N.  \label{sabun-disscs8}
\eeqa
\label{sabun-disscs101}\end{mathletters}This set of non-linear difference equations may be solved 
by identifying conserved quantities.
Let
\beq
P_{n} := \bar{\xi}^{S}_{n}\xi^{S}_{n-1},\qquad R_{n} := \bar{\xi}^{S}_{n}\xi^{S}_{n}, \label{def-pr}
\eeq
then it follows from Eqs.~(\ref{sabun-disscs101}) that
\begin{mathletters}\beqa
&& \frac { R_{n} - P_{n} }{ 1 + R_{n} } 
         =-i \varepsilon \frac{P_{n}}{ 1 + P_{n} } \qquad :1 \le n \le N-1, 
     \lab{const-eq1}\\ 
&& \frac { P_{n} - R_{n-1} }{ 1 + R_{n-1} } 
         =i \varepsilon \frac{P_{n}}{ 1 + P_{n} } \qquad :2 \le n \le N .
     \lab{const-eq2}
\eeqa
\end{mathletters}Putting these equations together, we find
\beqa
\frac{ 1 + P_{n} }{ 1 + R_{n-1} } = \frac{ 1 + P_{n} }{ 1 + R_{n} } 
                     \qquad :2 \le n \le N-1 .
\eeqa
If $1 + P_n$ vanished, 
the contribution of the stationary-action path to the amplitude in question
would vanish because of the factor $\ln (1 + P_n)$ in Eq.~(\ref{action-disscs}).
Hence we can assume that $1 + P_n \neq 0$. 
Consequently $R_n$ is a conserved quantity, whose value is to be denoted by $R$:\beqa
R_{n} = R \qquad :1 \le n \le N-1. \lab{constR}
\eeqa
Combining this with Eq.~(\ref{const-eq1}), 
we find that $P_n$ is also conserved and denote its value by $P$:
\beqa
 P_n = P := (1+i\varepsilon)R + \order(\varepsilon^2), \qquad 2 \le n \le N-1.   \lab{constP}
\eeqa     
(We can disregard the other root, 
which is equal to $-1 + \order(\varepsilon)$.)
With these results, 
the set of equations (\ref{sabun-disscs101}) reduces to
\begin{mathletters}\beqa
&& \xi^{S}_{n} - \xi^{S}_{n-1} = -i\varepsilon \xi^{S}_{n-1} + \order(\varepsilon^2)  \qquad :1\le n \le N-1,
   \lab{sabun-disscs9}\\
&& \bar{\xi}^{S}_{n} - \bar{\xi}^{S}_{n-1} = ~~i \varepsilon \bar{\xi}^{S}_{n}~~~ +  \order(\varepsilon^2)\qquad :2 \le n \le N.
   \lab{sabun-disscs10}
\eeqa\end{mathletters}Thus, to $\order(\varepsilon)$, the stationary-action path
expressed by $\xi^S$ and $\bar{\xi}^S$ is identical with 
that for the harmonic oscillator given by (\ref{sap-discs}).
 Accordingly
\beqa
R = \xi^{*}_{F}\xi_{I}e^{-iT} + \order(\varepsilon) ,\qquad P = \xi^{*}_{F}\xi_{I}e^{-iT+i\varepsilon} + \order(\varepsilon^{2}). \label{extra}
\eeqa
(We have illustrated how the stationary-action path may be found 
in the fully discrete form.
The result to the lowest order in $\varepsilon$ can also
 be found by going over to a differential equation
at the stage of Eqs.~(\ref{sabun-disscs101}).)
The result may be converted into $(\{\theta^S\}, \{\phi^S\})$ via 
\beq
      \xi^{S} = e^{i\phi^{S}}\tan \frac{\theta^{S}}{2},\qquad 
\bar{\xi}^{S} = e^{-i\phi^{S}}\tan \frac{\theta^{S}}{2},
\eeq
or equivalently into ${\bf n}^S$.
It is depicted in Fig.~3.

=== Fig. 3 ===
 
Again, the stationary-action path is not in general real.
Neither of them coincides with the classical path 
connecting ${\bf n}_I$ and ${\bf n}_F$ in time $T$; 
such a classical path would exist only in the special case (\ref{special-bc}).

\subsubsection{$\varepsilon$-term of the action}
If one regarded all the differences $|\xi_n - \xi_{n-1}|$ 
( $n = 1,2,\cdots,N$  with the convention $\xi_N := \xi_F$ and $\xi_0 := \xi_I$) as small in some sense and expanded the action to the second order in them,
one would obtain (the equality so found is to be denoted by $\sim$)
\begin{mathletters}\beqa
&& {\cal S}_{SCS}[\{\xi^* \}, \{\xi \}] 
\sim  {\cal S}_{SCS-\varepsilon 1}+{\cal S}_{SCS-\varepsilon 2}
    + {\cal S}_{SCS-c} +{\cal S}_{SCS-d} , \lab{separated-action-disscs}\\
&&  \frac{i}{\hbar}{\cal S}_{SCS-\varepsilon 1} := -S\sum_{n=1}^{N}\frac{1}{( 1 + |\xi_{n}|^{2} )^{2}}(\xi^{*}_{n}-\xi^{*}_{n-1})( \xi_{n} - \xi_{n-1} )   ,
\lab{1eterm-action-disscs}\\
&&  \frac{i}{\hbar}{\cal S}_{SCS-\varepsilon 2} := \frac{S}{2}\sum_{n=1}^{N}\frac{1}{( 1 + |\xi_{n}|^{2} )^{2}}[ \{\xi_{n}(\xi^{*}_{n}-\xi^{*}_{n-1})\}^{2} - \{ \xi^{*}_{n}( \xi_{n} - \xi_{n-1} ) \}^{2}] , 
\lab{2eterm-action-disscs}\\
&& \frac{i}{\hbar}{\cal S}_{SCS-c} := S \sum_{n=1}^{N}\frac{1}{ 1 + |\xi_{n}|^{2} }\{  \xi_{n}( \xi^{*}_{n} - \xi^{*}_{n-1}) - \xi^{*}_{n}(\xi_{n} - \xi_{n-1} )\}, \lab{cterm-action-disscs} \\
&& {\cal S}_{SCS-d} := -\varepsilon\sum_{n=1}^{N} {\cal H}( \xi^{*}_{n} , \xi_{n-1} ) . \lab{dterm-action-disscs}
\eeqa \label{actions-disscs}\end{mathletters}The four terms (\ref{actions-disscs}b-e) are to be called the "$\varepsilon 1$-term", "$\varepsilon 2$-term", the canonical term, and the dynamical term, respectively. In terms of $(\theta_{n},\phi_{n})$, they take the following forms:
\begin{mathletters}\beqa
\frac{i}{\hbar}{\cal S}_{SCS-\varepsilon 1} &\sim& \frac{i}{\hbar}\tilde{{\cal S}}_{SCS-\varepsilon 1}[\{\theta\},\{\phi\}]\nonumber\\ &:=& -\frac{S}{4}\sum_{n=1}^{N}\{ (\theta_{n} - \theta_{n-1})^{2} + (\phi_{n}-\phi_{n-1})^{2}(\sin\theta_{n})^{2} \} , \label{e1term-action-tp} \\
\frac{i}{\hbar}{\cal S}_{SCS-\varepsilon 2} &\sim& \frac{i}{\hbar}\tilde{{\cal S}}_{SCS-\varepsilon 2}[\{\theta\},\{\phi\}]\nonumber\\ &:=& -iS\sum_{n=1}^{N} (\theta_{n}-\theta_{n-1})(\phi_{n}-\phi_{n-1}) \tan \frac{\theta_{n}}{2}\left(\sin\frac{\theta_{n}}{2}\right)^{2}   ,
\label{e2term-action-tp}\\
\frac{i}{\hbar}{\cal S}_{SCS-c} &\sim& \frac{i}{\hbar}\tilde{{\cal S}}_{SCS-c}[\{\theta\},\{\phi\}]\nonumber\\ &:=& iS \sum_{n=1}^{N} \left\{ (\phi_{n}-\phi_{n-1})(\cos \theta_{n}-1) 
+ (\theta_{n}-\theta_{n-1})(\phi_{n}-\phi_{n-1}) \tan \frac{\theta_{n}}{2} \right\} .\label{cterm-action-tp}
\eeqa\label{actions-disscs-tp}\end{mathletters}Thus ${\cal S}_{SCS-\varepsilon 1}$ would resemble Klauder's $\epsilon$-term if the manipulation analogous to (\ref{keterm-action}) were correct. However, there is no reason to neglect ${\cal S}_{SCS-\varepsilon 2}$, which is also of the second order in $(\xi_{n}-\xi_{n-1})$. Eq.~(\ref{e2term-action-tp}) shows that it would give rise to 
\beqa
-iS \varepsilon \int_{0}^{T} dt \dot{\theta}(t)\dot{\phi}(t) \tan \frac{\theta(t)}{2}\left(\sin \frac{\theta(t)}{2}\right)^{2}
\eeqa
in addition to Klauder's $\epsilon$-term. It is seen from Eq.~(\ref{cterm-action-tp}) that a term of the same form emerges also from the canonical term, which is linear in $|\xi_{n}-\xi_{n-1}|$. This is due to the nonlinearity of the transformation (\ref{xi-phi-theta}).

\subsubsection{stationary action}
We now evaluate the contributions of each term of (\ref{actions-disscs}) separately
to the stationary action.
By an argument similar to that for the coherent-state case, it is shown that the $\varepsilon 1$-term does not contribute to the stationary action.
\beqa
{\cal S}_{SCS-\varepsilon 1}[\{\bar{\xi}^{S}\},\{\xi^{S}\}] = \order(\varepsilon).\label{contri-e1term}
\eeqa
In the same way, if the $\varepsilon 2 $-term is separated into three parts, the contribution from the intermediate times is obviously of $\order(\varepsilon)$, while the initial and final discontinuity respectively contributes
\beqa
\frac{S}{2(1+R)^{2}}(R-|\xi_{I}|^{2})^{2} + \order(\varepsilon),\qquad -\frac{S}{2(1+|\xi_{F}|)^{2}}(R-|\xi_{F}|^{2})^{2}+ \order(\varepsilon).
\lab{contribution-initial-final}
\eeqa
In contrast to the case of the $\varepsilon 1$-term, neither of these vanishes. 

The contribution of the canonical term can be separated into three parts as well. 
The contribution from the intermediate times gives
\beqa
iST\frac{2R}{1+R} + \order(\varepsilon),\lab{contribution-canonical-inter}
\eeqa
while the initial and final discontinuity respectively contributes
\beqa
\frac{S}{1+R}(R-|\xi_{I}|^{2}),\qquad \frac{S}{1+|\xi_{F}|^{2}}(R-|\xi_{F}|^{2}).
\lab{contribution-canonical-initial-final}
\eeqa
Finally, the contribution of the dynamical term, which contains the factor $\varepsilon$, comes from the intermediate times alone and gives
\beqa
iST\frac{1-R}{1+R} + \order(\varepsilon).\lab{contribution-dynamical}
\eeqa
The sum of (\ref{contribution-canonical-inter}) and (\ref{contribution-dynamical}) is equal to $iST$, which would reproduce the correct result (\ref{exact-scs-ta}) only in the special case (\ref{special-bc}); in this special case, the discontinuity terms (\ref{contribution-initial-final}) and (\ref{contribution-canonical-initial-final}) vanish as well.
However, in a generic case, the contribution of the discontinuity terms does not vanish, and the correct result is not reproduced even if one neglects the $\varepsilon 2$-term. The fluctuation integral can not remedy the result, either.

If one evaluated the stationary action by use of (\ref{actions-disscs-tp}), the discontinuity contributions from the $\varepsilon 1$-term do not vanish:
\beqa
\tilde{{\cal S}}_{SCS-\varepsilon 1}[\{ \theta^{S}\},\{\phi^{S}\}] &=& 
-\frac{S}{4}\Bigg[ \theta_{F}^{2}+\theta_{I}^{2} -4 \tan^{-1}\{\sqrt{R}\} \left( \theta_{F}+\theta_{I} -2\tan^{-1}\sqrt{R} \right) \nonumber \\ 
&& -(\sin\theta_{F})^{2}\left(\ln \frac{\sqrt{R}}{\tan \frac{\theta_{F}}{2}}\right)^{2} - \frac{4R}{(1+R)^{2}}\left(\ln \frac{\sqrt{R}}{\tan \frac{\theta_{I}}{2}}\right)^{2} \Bigg].
\eeqa
The apparent contradiction between this and (\ref{contri-e1term}) is caused by the interplay of the nonlinearity of (\ref{xi-phi-theta}) and the unwarranted negligence of the discontinuities at the initial and final times in writing
 (\ref{actions-disscs}) and (\ref{actions-disscs-tp}); ${\cal S}_{SCS-\varepsilon 1}[\{\bar{\xi}^{S}\},\{\xi^{S}\}]$ is not equivalent to $\tilde{{\cal S}}_{SCS-\varepsilon 1}[\{\theta^{S}\},\{\phi^{S}\}]$. 
A closely related ambiguity exists in the expression (\ref{2eterm-action-disscs}). If $|\xi_{n}-\xi_{n-1}|$ is small, the factor $ 1 + |\xi_{n}|^{2}$ could be replaced by $1+|\xi_{n-1}|^{2}$, for instance. The coresponding factor $1 + |\xi_{F}|^{2}$ in 
(\ref{contribution-initial-final}) would then be replaced by $1+R$, and so on.
 Hence the expansion in $|\xi_{n}-\xi_{n-1}|$ is not unique due to the nonlinearity and discontinuity. In any case, neither (\ref{actions-disscs}b-d) nor (\ref{actions-disscs-tp}) reproduces the correct result.

Therefore, we must go back to the original action (\ref{action-disscs}).
Substituting (\ref{def-pr}) into it, we find
\beqa
\frac{i}{\hbar} {\cal S}_{SCS}^{SAP} &:=& \frac{i}{\hbar} {\cal S}_{SCS} [ \{\bar{\xi}^{S}\} , \{\xi^{S}\}] \nonumber\\
 &=& S\ln \frac{(1+\xi^{*}_{F}\xi^{S}_{N-1})^{2}}{1+|\xi_{F}|^{2}}
+S \sum_{n=2}^{N-1}\left(2\ln \frac{1+P}{1+R} + i\varepsilon \frac{1-P}{1+P}\right) \nonumber \\
&&-2S\ln(1+R) 
+S\ln \frac{(1+\bar{\xi}^{S}_{1}\xi_{I})^{2}}{1+|\xi_{I}|^{2}} + \order(\varepsilon).
\lab{stationary-action-disscs}
\eeqa
The second term, where the value of $P$ to be used should be Eq. (\ref{constP})  correct up to $\order(\varepsilon)$, gives $ iST + \order(\varepsilon)$, while the final discontinuity (the first term) and the initial discontinuity (the last term) give  
\beqa
S\ln \frac{(1+R)^{4} }{( 1+ |\xi_{F}|^{2} )(1 + | \xi_{I} |^{2}) }+ \order(\varepsilon) ,\lab{exaxt-sa-disscs}
\eeqa
thereby  reproducing the complete amplitude (\ref{exact-scs-ta}).

Conclusion: If the action is expanded to the second order 
in the differences $|\xi_n - \xi_{n-1}|$ 
and those terms which would corresponding to Klauder's $\epsilon$-term are kept,
the resulting action does not reproduce the correct result (\ref{exact-scs-ta}). It is essential to respect
the "discontinuities at the initial and final times".

\subsubsection{fluctuations}
Let us evaluate the fluctuation integral and 
show that it reduces to unity.
We start from the proper discrete action (\ref{action-disscs}).
(If one started from (\ref{separated-action-disscs}), one would arrive at a non-sensical result 
as in the previous subsection.)
Decomposing the integration variables as (\ref{decomposed-xi-disscs}),
we find
\begin{mathletters}\beqa
&&{\cal S}_{SCS}[\{\xi^* \}, \{\xi \}] \simeq 
        {\cal S}_{SCS}^{SAP} + {\cal O}(\varepsilon)
         +  {\cal S}^{(2)}_{SCS }[\{\eta^* \}, \{\eta \}],
                 \lab{sa-fluc2-disscs}\\
&&{\cal S}^{(2)}_{SCS}[\{\eta^* \}, \{\eta \}] :=
 \sum_{n=1}^{N-1}\frac{i\hbar S}{(1+ \bar{\xi}^{S}_{n}\xi^{S}_{n})^{2}}\eta^{*}_{n}\eta_{n} 
+ \sum_{n=2}^{N-1}\left( \frac{-i\hbar S}{(1+ \bar{\xi}^{S}_{n}\xi^{S}_{n-1})^{2}} - \left.\frac{\partial^{2}{\cal H}(\xi^{*}_{n},\xi_{n-1})}{\partial \xi^{*}_{n}\partial\xi_{n-1}}\right|_{S}\right) \eta^{*}_{n}\eta_{n-1} \nonumber \\
&-&\varepsilon\sum_{n=1}^{N-1}\frac{S}{(1+ \bar{\xi}^{S}_{n}\xi^{S}_{n-1})^{2}}\left.\left\{\frac{\partial}{\partial \xi^{*}_{n}}\left( ( 1 + \xi^{*}_{n}\xi_{n-1})^{2} \frac{\partial {\cal H}(\xi^{*}_{n},\xi_{n-1})}{\partial \xi^{*}_{n}} \right)\right\}\right|_{S} (\eta^{*}_{n})^{2} \nonumber \\
&-&\varepsilon\sum_{n=2}^{N}\frac{S}{(1+ \bar{\xi}^{S}_{n}\xi^{S}_{n-1})^{2}}\left.\left\{\frac{\partial}{\partial \xi_{n-1}}\left(( 1 + \xi^{*}_{n}\xi_{n-1})^{2} \frac{\partial {\cal H}(\xi^{*}_{n},\xi_{n-1})}{\partial \xi_{n-1}} \right)\right\}\right|_{S} (\eta_{n-1})^{2}\Bigg] ,
                                \lab{fluc-action-disscs}
\eeqa
\end{mathletters}with the symbol $\simeq$ indicating equality 
up to the second order in $\eta_{n}$'s.
Accordingly the amplitude under consideration may be 
approximated as
\begin{mathletters}\beqa
\< \xi_{F} | e^{ -i\hat{H}T/\hbar  } | \xi_{I} \> &\simeq& \exp \left( \frac{i}{\hbar} {\cal S}_{SCS}^{SAP} \right)  {\cal K}^{(2)}_{SCS}(T) , \\
{\cal K}^{(2)}_{SCS} (T) &=& \lim_{ N \to \infty} \int\prod_{n=1}^{N-1}
\frac{2S+1}{2\pi i} \frac{ d \eta_{n}  d \eta^{*}_{n}  }{\{(1+\bar{\xi}^{S}_{n}+\eta^{*}_{n})(1+\xi^{S}_{n}+\eta_{n})\}^{2}} \nonumber\\
&&\times\exp \left( \frac{i}{\hbar} {\cal S}^{(2)}_{SCS} [\{\eta^{*}\},\{\eta\}] \right).
\lab{integral-fluc-disscs}
\eeqa
\end{mathletters}Since the integrand contains 
the Gaussian factor with exponent proportional to $-S\eta^{*}_{n}\eta_{n}$,
 the effective range of integration 
over $\eta_n$ is $|\eta_n| \lsim S^{-1/2}$.
Thus, $\eta_n$'s can legitimately be said to constitute 
a small fluctuation provided that $S \gg 1$.
 Accordingly the present stationary-action approximation is in fact an expansion with respect to $1/S$, 
in agreement with the intuition 
that a spin should behave "semi-classically" for large $S$. In the case of the spin under a constant magnetic field, the coefficients of the third and fourth term of (\ref{fluc-action-disscs}) vanish; ${\cal S}^{(2)}_{SCS}$ reduces to 
\begin{mathletters}\beqa
&&   \frac{i}{\hbar}{\cal S}^{(2)}_{SCS}[\{\eta^* \}, \{\eta \}] = -\frac{2S}{(1+R)^{2}} \sum_{n=1}^{N-1} 
\eta^{*}_{n}( \eta_{n} -  \alpha \eta_{n-1}) , \\ 
&& \alpha := (1-i\varepsilon)\left(\frac{1+R}{1+P}\right)^{2}, \qquad \eta_{0}:=0.
\eeqa
\end{mathletters}Hence, we make the formally same change of the integration variables as in (\ref{variable-change}) to find 
\beqa
{\cal K}^{(2)}_{SCS}(T) &=&  \lim_{N \to \infty}\int\prod_{n=1}^{N-1}
\frac{2S}{2\pi i}\frac{ d \eta^{'}_{n} d \eta^{'*}_{n} }{( 1 + R ) ^{2} } \exp \left( 
-\frac{2S}{ ( 1 + R )^{2} } \eta^{'*}_{n} \eta^{'}_{n} 
\right)\left\{1+\order\left(\frac{1}{S}\right)\right\}  \nonumber \\
&=&  1 + \order\left(\frac{1}{S}\right).
\eeqa
The complete amplitude (\ref{exact-scs-ta}) is thus recovered by treating DTSCSPI 
in the stationary-action approximation provided that $S \gg 1$
\cite{fn7}.


What would one find if one evaluated the fluctuation integrl starting from Klauder's action. Formal expansion as Eq. (\ref{expand-xiseta}) would give Eq. (\ref{expand-sapfluc}) with the subscript $CS$ replaced by $KSCS$, where
\beqa
S^{(2)}_{KSCS}[\eta^{*},\eta] &:=& \frac{i\hbar S}{(1+R)^{2}}\int_{0}^{T}dt\bigg[ \epsilon\bigg\{ \dot{\eta}^{*}(t)\dot{\eta}(t) - \frac{2R(1-2R)}{(1+R)^{2}}\eta^{*}(t)\eta(t)\nonumber\\
&&-\frac{2iR}{1+R}(\eta^{*}(t)\dot{\eta}(t)-\dot{\eta}^{*}(t)\eta(t))\bigg\}+\frac{4iR}{1+R}\eta^{*}(t)\eta(t) \nonumber\\
&&+ \eta^{*}(t)\dot{\eta}(t)-\dot{\eta}^{*}(t)\eta(t) + 2i\frac{1-R}{1+R} \eta^{*}(t)\eta(t) \bigg].
\eeqa
(Note that $R^{S}=R$.) One might adopt the discretization 
\beqa
S^{(2)}_{KSCS}[\eta^{*},\eta] &\to& {\cal S}^{(2)}_{KSCS}[\{\eta^{*}\},\{\eta\}]\nonumber\\
&:=& \frac{i\hbar S}{(1+R)^{2}}\sum_{n=1}^{N}
\bigg[ ( \eta^{*}_{n} - \eta^{*}_{n-1})(\eta_{n} - \eta_{n-1}) -\varepsilon^{2}\frac{2R(1-2R)}{(1+R)^{2}}\eta^{*}_{n}\eta_{n} \nonumber \\
 &-& \varepsilon \frac{2iR}{(1+R)^{2}}\{ \eta^{*}_{n}(\eta_{n}-\eta_{n-1})-(\eta^{*}_{n}-\eta^{*}_{n-1})\eta_{n} \}+ \varepsilon\frac{4iR}{1+R}\eta^{*}_{n}\eta_{n}\nonumber\\
&& + \{ \eta^{*}_{n}(\eta_{n}-\eta_{n-1})-(\eta^{*}_{n}-\eta^{*}_{n-1})\eta_{n}\} + 2i\varepsilon \frac{1-R}{1+R} \eta^{*}_{n}\eta_{n-1} \bigg], \label{flucint-app1}           
\eeqa
supplemented with 
\beqa
{\cal D}\eta{\cal D}\eta^{*} \to \prod_{n=1}^{N-1}\frac{2S}{2\pi i}\frac{d\eta_{n}d\eta^{*}_{n}}{(1+R)^{2}}. 
\eeqa
This would lead to a wrong result (see Appendix A):
\beqa
\int {\cal D}\eta{\cal D}\eta^{*} \exp\left(\frac{i}{\hbar}S^{(2)}_{KSCS}[\eta^{*},\eta]  \right) &\to& \lim_{N\to\infty}\int\prod_{n=1}^{N-1}\frac{2S}{2\pi i}
\frac{d\eta_{n}d\eta^{*}_{n}}{(1+R)^{2}}\exp\left(\frac{i}{\hbar}{\cal S}^{(2)}_{KSCS}[\{\eta^{*}\},\{\eta\}]\right)\nonumber\\&& = \exp\left\{-i\frac{(1+2R)R}{(1+R)^{2}}T\right\}. \label{flucint-app2}
\eeqa
If one neglected terms proportional to $\varepsilon$ except for the Hamiltonian term, the fluctuation integral would give unity. However, this negligence is {\em ad~hoc}. Otherwise, the fluctuation integral would not give unity. Other discretization schemes do not work either. This is the difficulty encountered by Klauder's $\epsilon$-prescription in the case of the spin-coherent-state path integral.

\acknowledgments
The authors are also grateful to T. Kashiwa for a valuable discussion on. 
This work has been supported by Grant-in-aid 
for Scientific Research C No.08640471 
and Grant-in-Aid for Scientific Research on Priority Areas No.271,
Japan Ministry of Education and Culture.   

\begin{appendix}
\section{calculation of fluctuation integral}\label{appenA}

In order to calculate the fluctuation integral (\ref{flucint-app2}), we may write ${\cal S}^{(2)}_{KSCS}$ defined by (\ref{flucint-app1}) in the following matrix expression: 
\beqa
\frac{i}{\hbar}{\cal S}^{(2)}_{KSCS} = -\frac{2S}{(1+R)^{2}}~ ^{t}{\bf \eta}^{*} {\cal M} {\bf \eta}
\eeqa
where the notation $ ^{t}{\bf \eta}^{*}$ and ${\bf \eta}$ is same as in Eq.~(\ref{matrix-exp}), and 
\beqa
{\cal M} :=
\left(
	\begin{array}{cccc}
a  &  -c    &        &         \\
-b & \ddots & \ddots &         \\
   & \ddots & \ddots &  -c     \\
   &        &  -b    &   a    
\end{array}\right),
\eeqa
with
\begin{mathletters}
\beqa
&& a:= 1 + i\varepsilon \frac{2R}{1+R} +\order(\varepsilon^{2}),  \\
&& b:= 1 - i\varepsilon \left( \frac{R}{(1+R)^{2}} + \frac{1-R}{1+R} \right), \\&& c:= i\varepsilon \frac{R}{(1+R)^{2}}.
\eeqa
\end{mathletters}The determinant of ${\cal M}$ is obtained as 
\beqa
\det {\cal M} = \frac{\lambda_{+}^{N-2}(a^{2}-bc-\lambda_{-}a)-\lambda_{-}^{N-2}(a^{2}-bc-\lambda_{+}a)}{\lambda_{+}-\lambda_{-}},
\eeqa
where
\beqa
&& \lambda_{+} := \frac{a+\sqrt{a^{2}-4bc}}{2} = 1 + i\varepsilon \frac{(1+2R)R}{(1+R)^{2}} + \order(\varepsilon^{2}), \\
&& \lambda_{-} := \frac{a-\sqrt{a^{2}-4bc}}{2} = i\varepsilon \frac{R}{(1+R)^{2}} + \order(\varepsilon^{2}).
\eeqa
Since $\lambda_{-}$ is $\order({\varepsilon})$, we can the $\det{\cal M}$ in the limit $N \to \infty$ to
\beqa
\det{\cal M} = \lambda^{N}_{+}\{ 1 +\order(\varepsilon) \} = \exp
\left\{  i\frac{(1+2R)R}{(1+R)^{2}}T  \right\} + \order(\varepsilon).
\eeqa
Hence, we obtain the result (\ref{flucint-app2}).

\end{appendix}

\newpage
\pagestyle{empty}
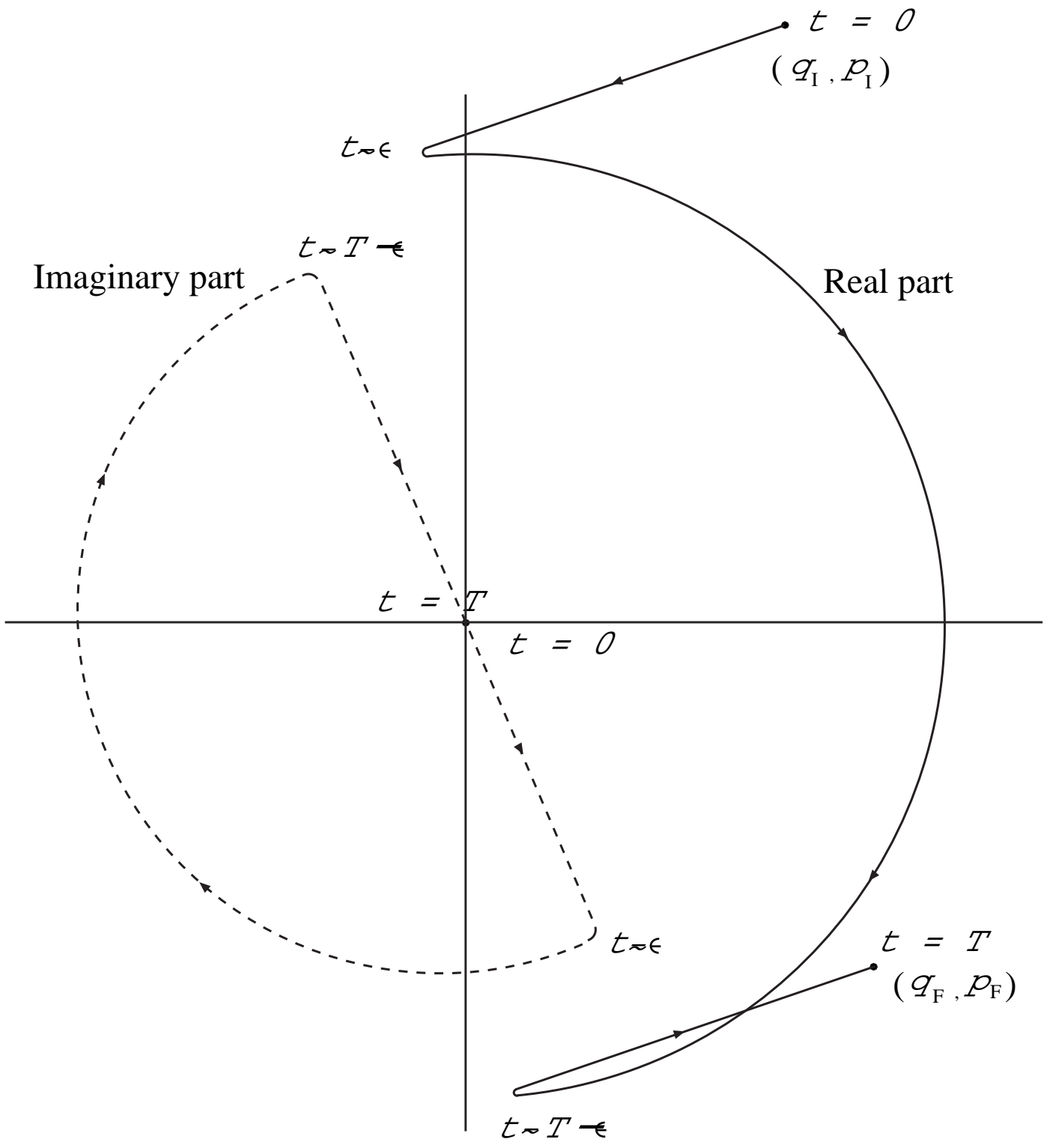
\begin{figure}
\caption{Klauder's stationary-action path for the harmonic oscillator
depicted in the phase space.
The real and imaginary parts are drawn with a solid curve and a dashed curve, respectively.}
\label{fig.1}
\end{figure}

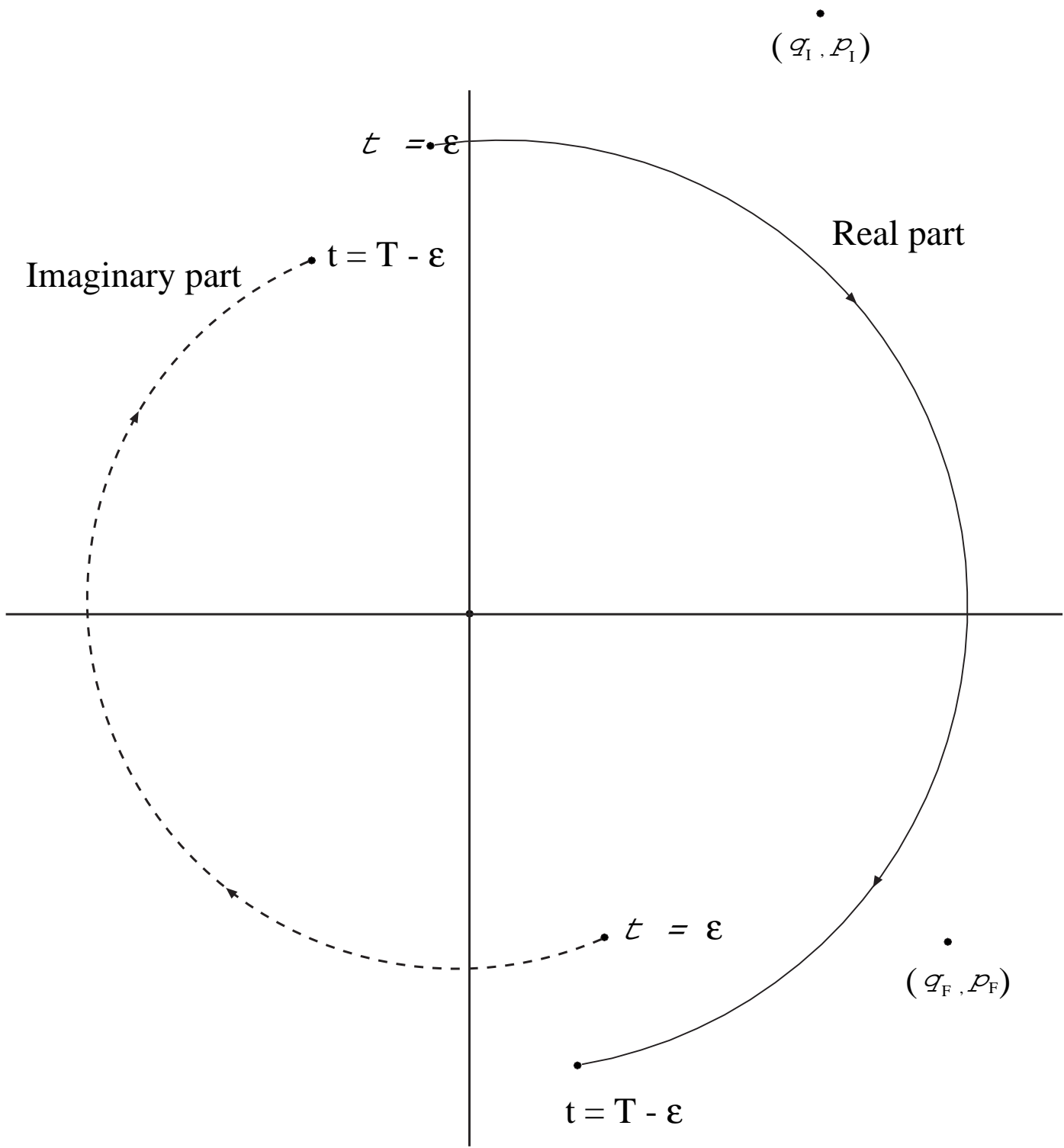
\begin{figure}
\caption{The stationary-action path for the harmonic oscillator
depicted in the phase space.
The real and imaginary parts are drawn with a solid curve and a dashed curve, respectively.}
\label{fig.2}
\end{figure}

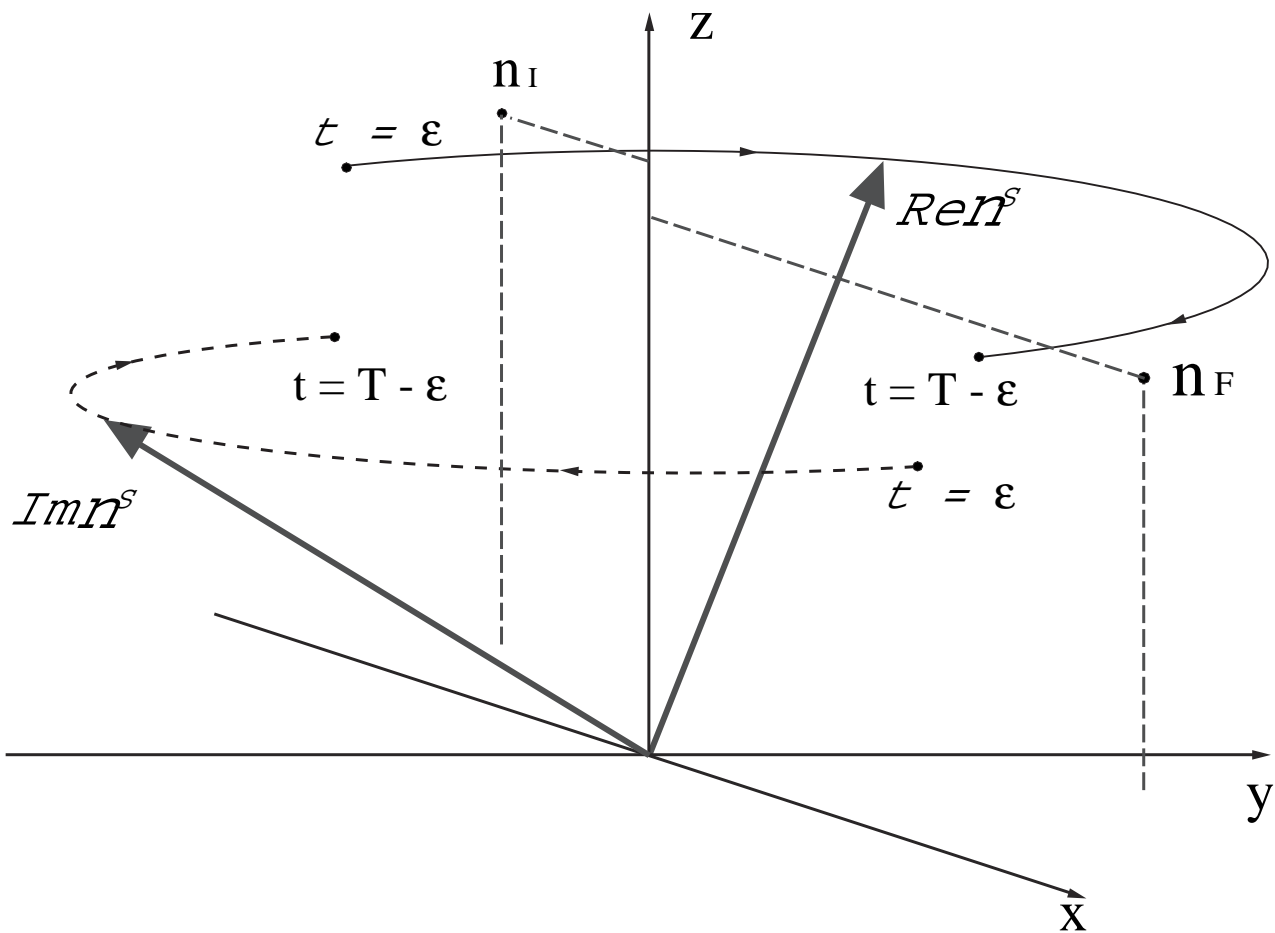
\begin{figure}
\caption{The stationary-action path for the spin 
under a constant magnetic field depicted in terms of ${\bf n}^{\it S}$.
The real and imaginary parts of ${\bf n}^{\it S}$
is drawn with a solid curve and a dashed curve, 
respectively.}
\label{fig.3}
\end{figure}


\begin{references}



\bibitem{Feynman}
R. P. Feynman and A. R. Hibbs, "Quantum Mechanics and Path Integrals," McGraw-Hill, New York, 1965.


\bibitem{Schulman}
L. S. Schulman, "Techniques and Applications of Path Integration," Wiley, New York, 1981.

\bibitem{Kleinert}
 H. Kleinert, "Path integrals in quantum mechanics, statistics, and polymer physics," 2nd ed., World Scientific, Singapore, 1995.


\bibitem{Swanson}
Mark S. Swanson, "Path Integrals and Quantum Processes," Academic Press, New York, 1992.





\bibitem{Kuratsuji}
H. Kuratsuji and Y. Mizobuchi, {\em J. Math. Phys.} {\bf 22}, (1981), 757.


\bibitem{Fradkin}
E. Fradkin, "Field Theories of Condensed Matter Systems," Chapter~5, Addison-Wesley, 1991. 

\bibitem{Chudnovsky}
E. M. Chudnovsky and L. Gunther, {\em Phys. Rev. Lett.} {\bf 60}, (1988), 661.


\bibitem{Braun-Loss}
H-B. Braun and D. Loss, {\em Phys. Rev. B} {\bf 53}, (1996), 3237.




\bibitem{Klauder}
J. R. Klauder, {\em Phys. Rev. D} {\bf 19}, (1979), 2349.



\bibitem{Funahashi1}
K. Funahashi, T. Kashiwa, S. Nima, and S. Sakoda, {\em Nucl. Phys.} {\bf B} 453 (1995), 508.


\bibitem{Funahashi2}
K. Funahashi, T. Kashiwa, and S. Sakoda, {\em J. Math. Phys.} {\bf 36}, (1995), 3232.


\bibitem{Schilling}
R. Schilling, in "Quantum Tunneling of Magnetization," Proceedings of the NATO workshop, Chichilianne, France, 1994, edited by L. Gunther and B. Barbara (Kluwer Academic, Norwell, MA, 1995), p.~59.


\bibitem{Solari}
H. G. Solari, {\em J. Math. Phys.} {\bf 28}, (1987), 1097.





\bibitem{fn1}
A partition function involves only those stationary-action paths
obeying the periodic boundary condition. 
Evaluation of transition amplitudes, on the other hand, requires
those with a generic boundary condition (see Sec. V).
In this sense the calculation performed in Sec. V is an extension
of that in Ref.\cite{{Funahashi1},{Funahashi2}}



\bibitem{Radcliffe}
J. M. Radcliffe, {\em J. Phys. A} {\bf 4}, (1971), 313.


\bibitem{fn2}
Except for the case that $T/\pi$ is an integer;
this is the well-known "caustic case" for a harmonic oscillator. 



\bibitem{Schul}
An alternative boundary condition $p^{S}(0) = p_{I},~q^{S}(0) = q_{I}$ is chosen at p.~248 of \cite{Schulman}. But there is no justification for this choice. 


\bibitem{Itzykson}
C. Itzykson and J-B. Zuber, "Quantum Field Theory," p.~ 438, McGraw-Hill, New York, 1980.

\bibitem{independent}
This is so in spite of the fact that $\bar{\xi}^{S}(t)$ is not necessarily a complex conjugate of $\xi(t)$. This property holds since the integration contours can be distorted into the complex planes independently for $p(t)$ and $q(t)$.


\bibitem{fn4}
This is not a prescription proposed by Klauder,
who did not address himself to the question of fluctuations
around his stationary-action path.

\bibitem{fn5}
In view of (\ref{exact-cs-ta}) and (\ref{sa-kctcs}), integration over fluctuations should give unity. 

\bibitem{fn1-1}
This point has not been mentioned explicitly standerd text books \cite{{Feynman},{Schulman},{Kleinert},{Swanson}}.




\bibitem{fn6}
This is merely a matter of convenience to render equations compact. No notion of "boundary condition" on the integration variables is involved.


\bibitem{1}
This is allowed by the reason mentioned at \cite{independent}. 

\bibitem{fn7}
See Ref.\cite{Funahashi1} for the reason why the stationary-action approximation gives the exact result in the case of a single spin under a constant magnetic field.

\end{references}
\end{document}